\documentclass[conference]{IEEEtran}
\IEEEoverridecommandlockouts
\usepackage{cite}
\usepackage{amsmath,amssymb,amsfonts}
\usepackage{algorithmic}
\usepackage{graphicx}
\usepackage{textcomp}
\usepackage{xcolor}
\usepackage{caption} 
\usepackage{multirow}
\usepackage{makecell}
\usepackage{booktabs}
\usepackage{subfig}
\usepackage{subcaption}
\def\BibTeX{{\rm B\kern-.05em{\sc i\kern-.025em b}\kern-.08em
    T\kern-.1667em\lower.7ex\hbox{E}\kern-.125emX}}
\usepackage{url}
\newcommand{\methodname}{scUnified}

\begin{document}

\title{scUnified: An AI-Ready Standardized Resource for Single-Cell RNA Sequencing Analysis
% scUnity-AI: An AI-Ready Single-Cell RNA Sequencing Resource with Standardized Processing for Multi-Task Computational Analyses
}

% \author{\IEEEauthorblockN{
% Anonymous
% % \thanks{This}
% }}

% \IEEEauthorblockA{\textit{$^1$Computer Network Information Center, Chinese Academy of Sciences, Beijing, China} \\
% \textit{$^2$University of Chinese Academy of Sciences, Beijing, China} \\
% \{zyc\}@cnic.cn
% }}

% \title{Conference Paper Title*\\
% {\footnotesize \textsuperscript{*}Note: Sub-titles are not captured for https://ieeexplore.ieee.org  and
% should not be used}
% \thanks{Identify applicable funding agency here. If none, delete this.}
% }

\author{\IEEEauthorblockN{Ping Xu$^{1,3}$, Zaitian Wang$^{1,3}$, Zhirui Wang$^{2,3}$, Pengjiang Li$^{1,3}$, Ran Zhang$^{1,3}$, Gaoyang Li$^{1,3}$, \\ Hanyu Xie$^{4}$, Jiajia Wang$^{1,3}$, Yuanchun Zhou$^{1,2,3}$, Pengfei Wang$^{1,2,3,*}$
\thanks{$^*$ Corresponding author.}
% \thanks{The resource is available at the link: \url{https://anonymous.4open.science/r/scBenckmark-180E}. }
\thanks{The resource is publicly available at the link: \protect\url{https://github.com/XPgogogo/scUnity-AI}.}
\thanks{This work was supported by the National Natural Science Foundation of China (Grant No. 92470204 and 62406306)and the National Key Research and Development Program of China Grant (No. 2024YFF0729201).}
}
\IEEEauthorblockA{\textit{$^1$Computer Network Information Center, Chinese Academy of Sciences, Beijing, China} \\
\textit{$^2$Hangzhou Institute for Advanced Study,University of Chinese Academy of Sciences, Hangzhou, China}\\
\textit{$^3$University of Chinese Academy of Sciences, Beijing, China} \\
\textit{$^4$Department of Computer Science, Columbia University, New York, USA} \\
% \{xuping,pjli,zhangran,zyc\}@cnic.cn, wangzaitian23@mails.ucas.ac.cn, wpf2106@gmail.com
% }}
xuping0098@gmail.com, wpf2106@gmail.com}}

\maketitle

\begin{abstract}
% Single-cell RNA sequencing (scRNA-seq) has become a powerful tool for dissecting cellular heterogeneity, yet its widespread application is hindered by the lack of standardized, analysis-ready datasets. Current resources often differ in formats, preprocessing workflows, and annotation strategies, which complicates model training, systematic evaluation, and reproducibility across studies. While many computational methods have been proposed for clustering, annotation, and trajectory inference, their benchmarking and deployment remain limited by data inconsistency.  
% Here, we present~\methodname, an AI-ready single-cell RNA-seq resource that consolidates 13 high-quality datasets spanning two species (human and mouse) and nine tissue types. All datasets undergo standardized quality control and preprocessing, and are stored in a uniform format to enable direct application in diverse computational analyses without additional data cleaning. We further demonstrate the utility of~\methodname~in representative downstream tasks, establishing a reproducible foundation for AI-driven method development and systematic evaluation across datasets.

Single-cell RNA sequencing (scRNA-seq) technology enables systematic delineation of cellular states and interactions, providing crucial insights into cellular heterogeneity.
Building on this potential, numerous computational methods have been developed for tasks such as cell clustering, cell type annotation, and marker gene identification.
To fully assess and compare these methods, standardized, analysis-ready datasets are essential.
However, such datasets remain scarce, and variations in data formats, preprocessing workflows, and annotation strategies hinder reproducibility and complicate systematic evaluation of existing methods.
To address these challenges, we present~\methodname, an AI-ready standardized resource for single-cell RNA sequencing data that consolidates 13 high-quality datasets spanning two species (human and mouse) and nine tissue types.
All datasets undergo standardized quality control and preprocessing and are stored in a uniform format to enable direct application in diverse computational analyses without additional data cleaning. 
We further demonstrate the utility of~\methodname~through experimental analyses of representative biological tasks, providing a reproducible foundation for the standardized evaluation of computational methods on a unified dataset.

\end{abstract}


\section{Introduction}
With the rapid advancement of single-cell RNA sequencing (scRNA-seq) technologies, it is now possible to characterize complex cellular populations and their functional states at unprecedented resolution~\cite{shapiro2013single}. 
Although scRNA-seq data are inherently high-dimensional, sparse, and subject to substantial technical noise, they capture rich biological information that provides a robust foundation for studying cellular heterogeneity and elucidating disease mechanisms~\cite{kiselev2019challenges, petegrosso2020machine, menon2018clustering, xu2024sccdcg}. 
Such data enable a wide array of computational analyses, including cell clustering, cell type annotation, trajectory inference, gene regulatory network reconstruction, and so on~\cite{wang2025sccompass, mereu2020benchmarking}.
% However, data generated across different experiments, sequencing platforms, and batches exhibit substantial heterogeneity in formats, preprocessing workflows, and annotation strategies, thereby posing significant challenges for downstream computational analyses~\cite{wang2025sccompass}.

Recent years have witnessed the development of diverse computational strategies for scRNA-seq data, spanning traditional statistical modeling, machine learning, deep learning, and foundation models informed by biological priors to cope with the complexity and noise inherent in single-cell datasets~\cite{zhang2023review, krzak2019benchmark, xu2025scsiameseclu, xu2025soft, qi2020clustering, zhai2023realistic}. 
Representative methods include graph-based clustering algorithms such as Louvain and Leiden~\cite{stuart2019comprehensive}, probabilistic generative models like scVI~\cite{lopez2018deep}, deep clustering frameworks such as scCDCG~\cite{xu2024sccdcg}, and large-scale foundation models including scGPT~\cite{cui2024scgpt} and GeneCompass~\cite{yang2024genecompass}.

Despite the rapid progress and methodological diversity, these computational models face significant obstacles in rigorous evaluation and comparison~\cite{brooks2024challenges, yu2022benchmarking, dai2022scimc, wang2022comparison}. 
Specifically, the lack of standardized, high-quality datasets limits reproducibility and hinders fair benchmarking. 
Three main challenges can be identified. 
(i) Unrigorous cluster number setting: In single-cell clustering benchmarks, the number of annotated cell types is often directly used as the number of clusters. This practice is not always biologically justified and may introduce biases in performance evaluation.
(ii) Inconsistent data standards leading to unfair evaluation: Significant variations exist across datasets in terms of format, preprocessing workflows, and annotation quality. These inconsistencies not only hinder model training across studies but also limit the ability to perform fair and reproducible comparisons of multiple methods on the same dataset.
(iii) Limited availability of multi-task datasets: Few existing single-cell datasets can simultaneously support diverse downstream biological analyses, such as clustering, cell-type annotation, and marker gene identification. This limitation restricts the scope of systematic benchmarking and constrains the advancement of AI-driven single-cell research.

\begin{figure*}[!h]
    \centering
    \includegraphics[width=1\linewidth]{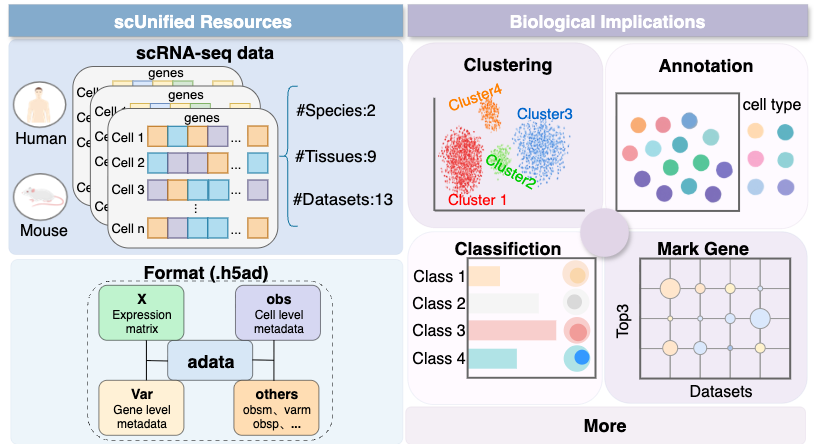}
    \caption{Overview of~\methodname: standardized single-cell RNA sequencing datasets across species and tissues, supporting AI-driven biological research and discovery.}
    \label{fig:framework}
\end{figure*}

% To address these challenges,
To cope with the aforementioned issues, we present \textbf{\methodname}, an \textbf{AI-ready standardized resource for single-cell RNA sequencing analysis}. 
\methodname~integrates 13 high-quality publicly available datasets covering two species and nine tissue types, with consistent quality control, preprocessing, and multi-level annotations provided in the \texttt{.h5ad} format to ensure compatibility with widely used single-cell analysis frameworks. 
By providing analysis-ready data, \methodname~eliminates the need for additional data cleaning or format conversion, offering a standardized and reliable resource that facilitates reproducible evaluation of computational methods across diverse models and tasks.
Our principal contributions are summarized as follows:
\begin{itemize}
\item comprehensive collection and systematic curation of high-quality scRNA-seq datasets with uniform quality control and preprocessing;
\item provision of standardized, analysis-ready data format to support a broad spectrum of biological tasks, including clustering, cell type annotation, marker gene identification, and beyond;
\item validation of dataset utility through representative biological case studies, establishing a reproducible foundation for method development, fair model comparison, and AI-driven discovery in single-cell research.
\end{itemize}

% 随着单细胞 RNA 测序（scRNA-seq）技术的快速发展，研究者能够以前所未有的分辨率刻画复杂的细胞群体及其功能状态。这类数据虽然具有高维度、稀疏性和较高的技术噪声，但仍蕴含丰富的生物学信号，为单细胞研究和疾病机制解析提供了坚实的数据基础。然而，不同实验、测序平台和批次生成的 scRNA-seq 数据往往存在显著的异构性和分散性，在数据格式、预处理流程及注释策略等方面缺乏统一标准，给后续的计算分析带来了较大挑战。

% 目前，scRNA-seq 的计算分析方法涵盖聚类分析、细胞类型分类与注释、轨迹推断、细胞状态预测等多类任务，并基于传统统计模型、机器学习、深度学习、图神经网络及生物学先验驱动等多种范式不断发展。尽管这些方法在性能与创新性方面取得了显著进展，但缺乏统一、标准化的数据集资源，使得模型训练、性能评估以及跨数据集的公平比较面临挑战。这种不足不仅妨碍了方法在不同任务与应用场景中的系统性检验，也削弱了研究成果的可复现性和跨领域推广能力。

% 针对上述问题，本工作构建了一个 AI-Ready 的 scRNA-seq 数据集资源，覆盖多物种、多组织和多细胞类型，并经过严格的质量控制与统一的预处理，采用标准化存储格式以支持多类单细胞分析任务。该资源可直接用于方法开发、模型训练与跨数据集评估，无需额外的数据清洗与格式转换。我们的主要贡献包括：
% （1）系统收集和整理多个高质量公开 scRNA-seq 数据集，并进行一致性的质控与预处理；
% （2）提供统一的数据格式，便于直接调用于各类分析任务；
% （3）通过案例分析展示数据集在多种任务场景下的适用性，为后续 AI 方法研究提供可复现的高质量数据基础。

% \input{2_relatedwork}
\section{\methodname}

% \begin{figure*}[!ht]
%     \centering
%     \includegraphics[width=0.8\linewidth]{image//framework/framework.png}
%     \caption{Overview of~\methodname: standardized scRNA-seq datasets across species and tissues, supporting AI-driven biological discovery.}
%     \label{fig:framework}
% \end{figure*}

We present~\textbf{\methodname}, an \textbf{AI-ready standardized resource for single-cell RNA sequencing analysis}, offering uniformly processed datasets that facilitate fair evaluation of diverse computational models on a consistent and reproducible dataset collection.
As depicted in Fig.~\ref{fig:framework},~\methodname~provides a standardized workflow that integrates and curates 13 publicly available single-cell RNA sequencing datasets spanning two species and nine tissue types.
All datasets undergo uniform quality control, preprocessing, and multi-level annotation, stored in the \texttt{.h5ad} format to ensure seamless compatibility with widely adopted single-cell analysis frameworks. 
This standardized resource enables direct application to a wide range of computational tasks, including core analyses such as cell clustering, cell type annotation, cell type classification, and marker gene identification, and supports fair evaluation of multiple models on the same dataset or a single model across multiple datasets, without the need for additional curation or format conversion.
By unifying diverse, high-quality scRNA-seq datasets into a consistent and accessible framework,~\methodname~provides a reproducible foundation for AI-driven single-cell research, reducing technical barriers, facilitating method development and training, and supporting rigorous systematic evaluation across biological contexts.

\section{Dataset}

\begin{table*}[!t]
    \centering
    % \scriptsize
    \small
   % \vspace{-5mm}
    \setlength{\tabcolsep}{4pt}
    \renewcommand{\arraystretch}{1}
    \begin{tabular}{llccccccc}
    \toprule
    \textbf{Species} & \textbf{Dataset Name} & \textbf{\#Cell} & \textbf{\#Gene} & \textbf{\#Cluster} & \textbf{Organ} & \textbf{Seq. Method} & \textbf{Sparsity (\%)} & \textbf{Ref.} \\
    \midrule
    \multirow{8}{*}{\textbf{\makecell{Human}}} 
    & Mauro Pancreas & 2,122 & 19,046 & 9 & Pancreas & CEL-seq2 & 73.02 & \cite{muraro2016single} \\
    & Sonya Liver & 8,444 & 4,999 & 11 & Liver & 10X Genomics & 90.77 & \cite{macparland2018single} \\
    & Sapiens Liver & 2,152 & \textbf{61,759} & 15 & Liver & Smart-seq2 & 95.42 & \cite{tabula2020single} \\
    & Sapiens Ear Crista Ampullaris & 2,357 & \textbf{61,759} & 7 & Ear & Smart-seq2 & 93.59 & \cite{tabula2020single} \\
    & Sapiens Ear Utricle & 611 & \textbf{61,759} & 5 & Ear & Smart-seq2 & 93.75 & \cite{tabula2020single} \\
    & Sapiens Lung & 6,530 & \textbf{61,759} & 25 & Lung & Smart-seq2 & 93.88 & \cite{tabula2020single} \\
    & Sapiens Testis & 7,494 & \textbf{61,759} & 8 & Testis & Smart-seq2 & 93.91 & \cite{tabula2020single} \\
    & Sapiens Trachea & \textbf{22,592} & \textbf{61,759} & 20 & Trachea & Smart-seq2 & 94.73 & \cite{tabula2020single} \\
    \midrule
    \multirow{5}{*}{\textbf{\makecell{Mouse}  }} 
    & Muris Limb Muscle & 3,855 & 21,609 & 6 & Limb Muscle & Smart-seq2 & 91.38 & \cite{tabula2020single} \\
    & Muris Brain & 13,417 & 21,609 & 2 & Brain & Smart-seq2 & 91.83 & \cite{tabula2020single} \\
    & Muris Kidney & 1,817 & 21,609 & 9 & Kidney & Smart-seq2 & 92.25 & \cite{tabula2020single} \\
    & Muris Liver & 2,859 & 21,609 & 11 & Liver & Smart-seq2 & 88.20 & \cite{tabula2020single} \\
    & Muris Lung & 5,167 & 21,609 & 25 & Lung & Smart-seq2 & 89.90 & \cite{tabula2020single} \\
    \bottomrule
    \end{tabular}
    \caption{Details of 13 selected single-cell gene expression datasets. Large datasets (with \( > 20000\) cells) and high-dimensional datasets (with \( > 60000\) genes) are highlighted with bold fonts. } 
    \label{tab:selected_dataset}
\end{table*}

\subsection{Dataset Coverage}
\label{app:appendix_dataset}

% \begin{figure*}[!t]
% \centering

% \subfloat[Distribution of human tissues]{
% \includegraphics[width=0.23\textwidth,keepaspectratio]{image/data/WS_Human_Organ.pdf}
% \label{subfig:WS_Human_Organ}
% }
% \subfloat[Distribution of mouse tissues]{
% \includegraphics[width=0.23\textwidth,keepaspectratio]{image/data/WS_Mouse_Organ.pdf}
% \label{subfig:WS_Mouse_Organ}
% }
% \subfloat[Number of Cells]{
% \includegraphics[width=0.23\textwidth,keepaspectratio]{image/data/num_cells_histogram.pdf}
% \label{subfig:cell}
% }
% \\
% \subfloat[Number of Genes]{
% \includegraphics[width=0.23\textwidth,keepaspectratio]{image/data/num_genes_histogram.pdf}
% \label{subfig:gene}
% }
% % \\
% \subfloat[Number of Clusters]{
% \includegraphics[width=0.23\textwidth,keepaspectratio]{image/data/num_clusters_histogram.pdf}
% \label{subfig:cluster}
% }
% \subfloat[Sparsity (\%)]{
% \includegraphics[width=0.23\textwidth,keepaspectratio]{image/data/sparsity_histogram.pdf}
% \label{subfig:sparsity}
% }
% % \\
% % % \vspace{-4mm}
% % \subfloat[Distribution of Sample Numbers in Human and Mouse Data]{
% % \includegraphics[width=0.3\textwidth,keepaspectratio]{image/data/bubble.png}
% % \label{subfig:organ}
% % }
% % \vspace{-3mm}
% \caption{Dataset distributions. (a) to (d) show dataset distributions by cell count, gene number, clusters, and sparsity.
% % , while (e) displays the distribution of samples between human and mouse datasets.
% % covering 8 human and 14 mouse types.
% }
% \label{fig:Data_Distribution}
% \end{figure*} 

\begin{figure*}[htbp]
    \centering
    \begin{minipage}[t]{0.3\textwidth}
        \centering
        \subfloat[Distribution of human tissues]{
        \includegraphics[width=\textwidth,height=0.18\textheight,keepaspectratio]{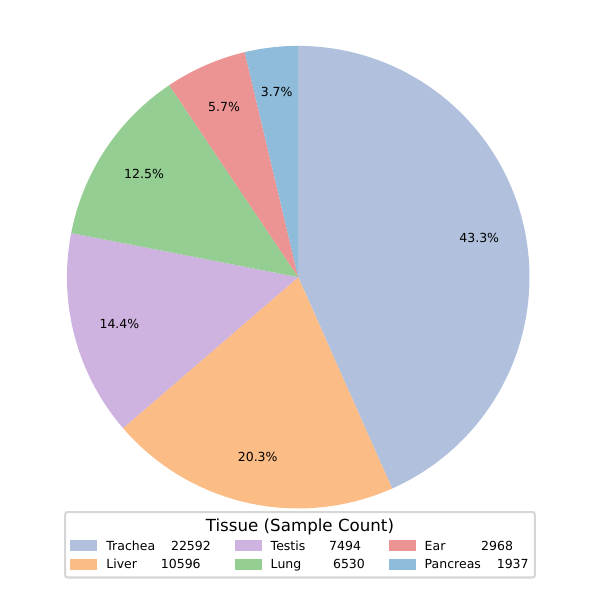}
        \label{subfig:WS_Human_Organ}
        }
        \\
        \subfloat[Distribution of mouse tissues]{
        \includegraphics[width=\textwidth,height=0.18\textheight,keepaspectratio]{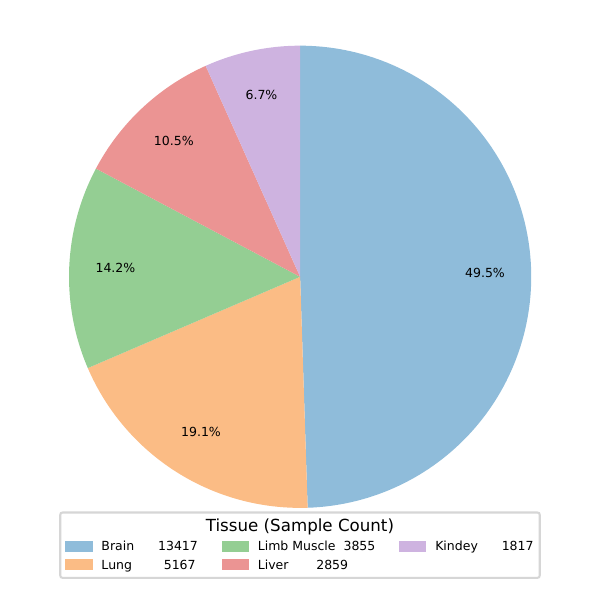}
        \label{subfig:WS_Mouse_Organ}
        }
        \caption{Data distribution of sample numbers in human and mouse data}
    \end{minipage}
    \hspace{4pt}
    \begin{minipage}[t]{0.5\textwidth}
        \centering
        \subfloat[Number of Cells]{% First square
            \includegraphics[width=0.45\linewidth]{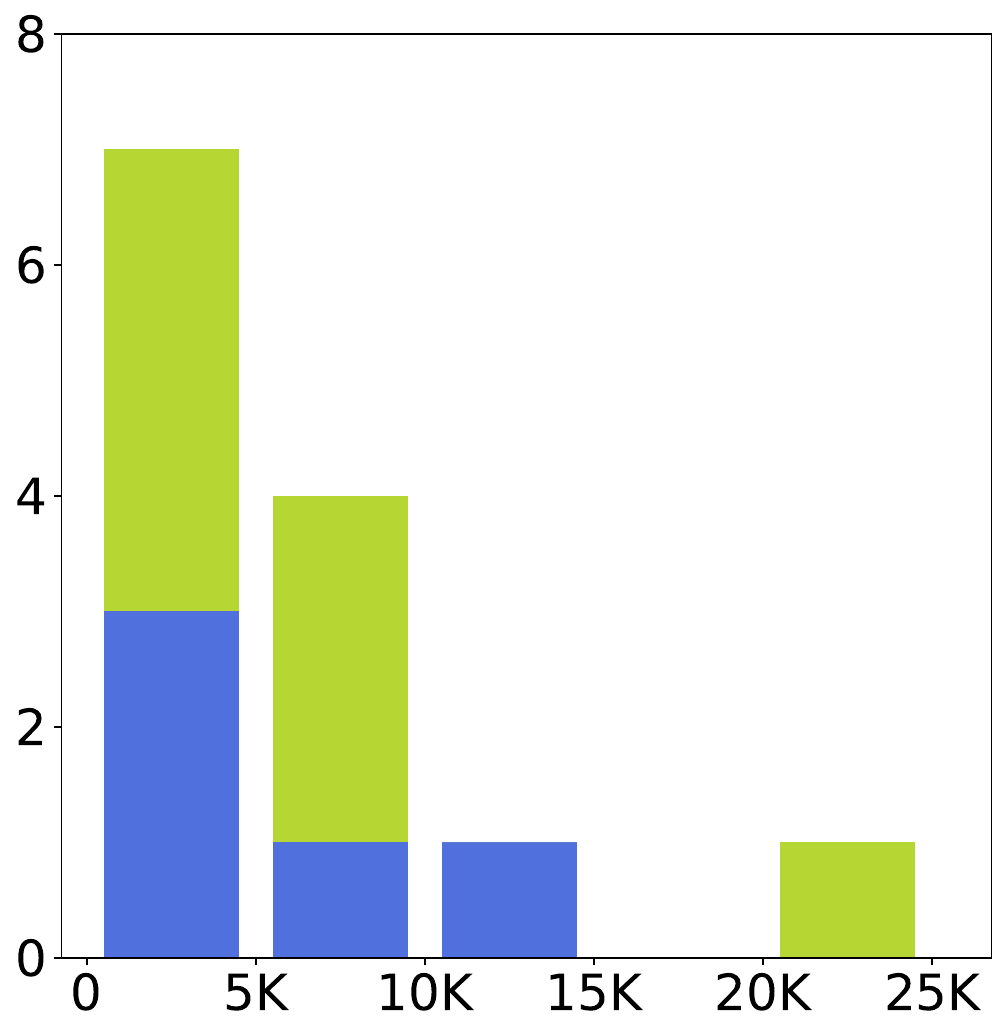}
            \label{subfig:cell}
        }
        \subfloat[Number of Genes]{% Second square
            \includegraphics[width=0.45\linewidth]{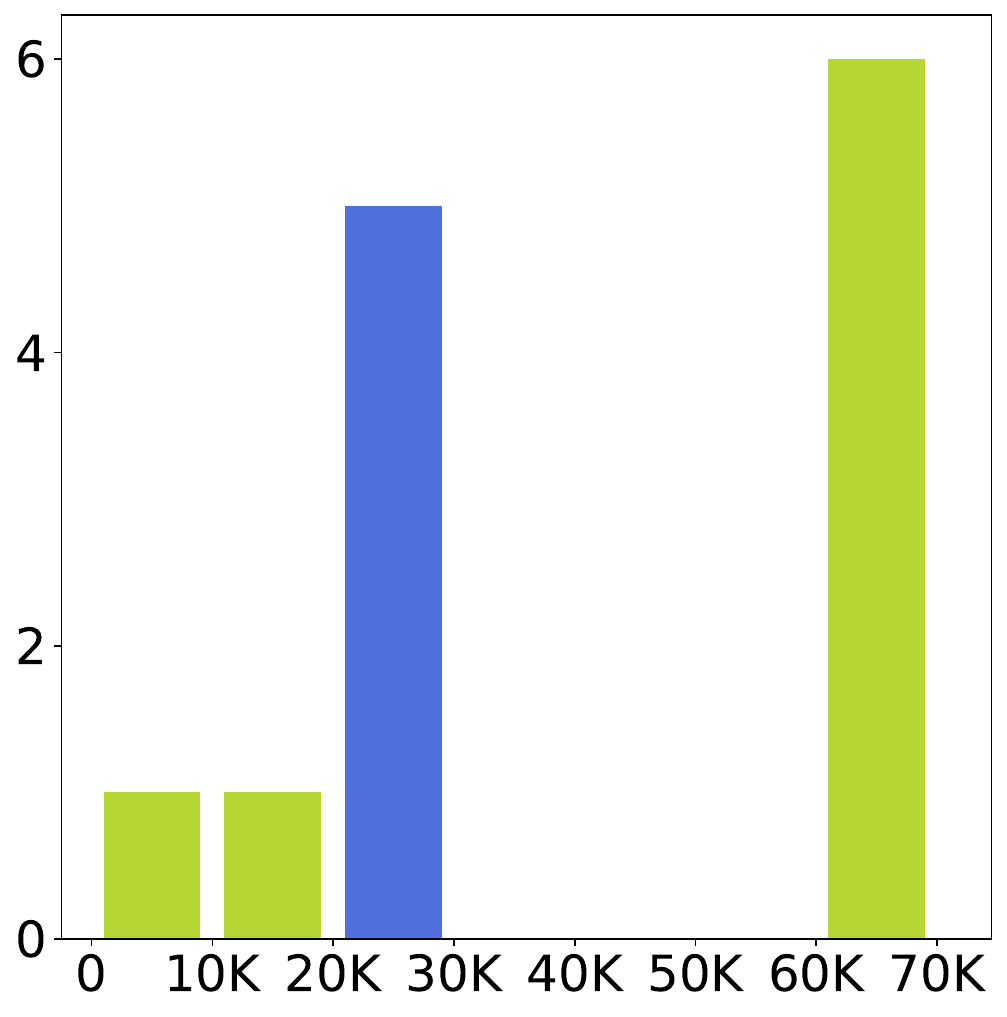}
            \label{subfig:gene}
        }
        \\
        \vspace{2pt}
        \subfloat[Number of Clusters]{% With caption
            \includegraphics[width=0.45\linewidth]{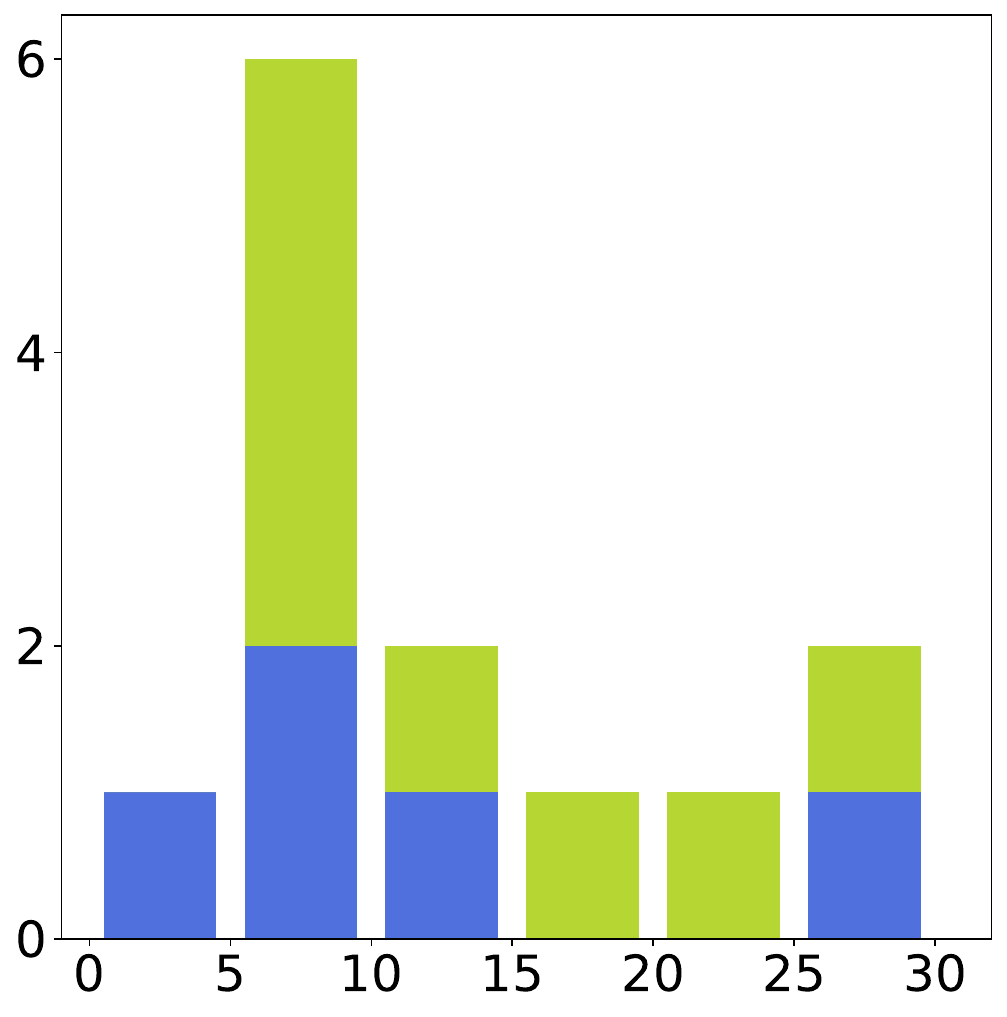}
            \label{subfig:cluster}
        }
        \subfloat[Sparsity (\%)]{% With caption
            \includegraphics[width=0.46\linewidth]{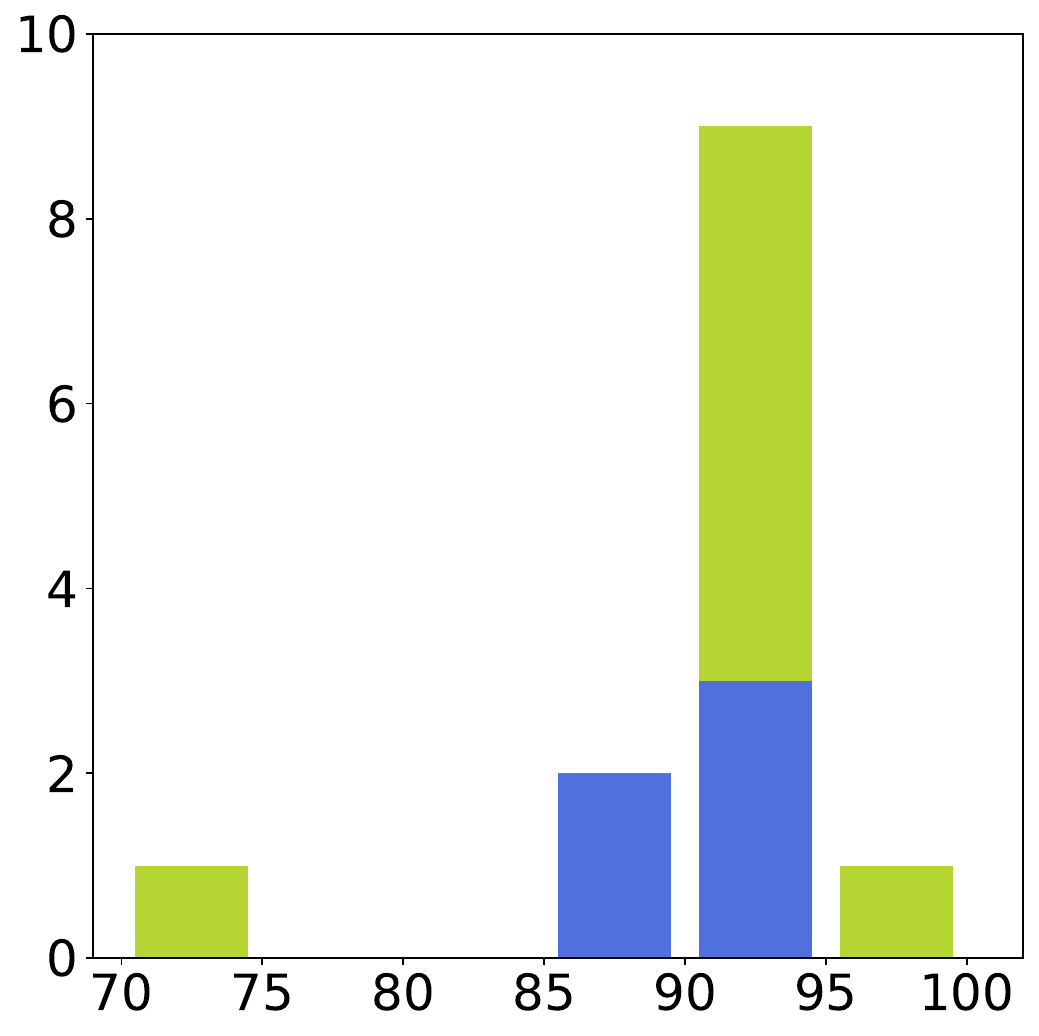}
            \label{subfig:sparsity}
        }
        \caption{Dataset distributions by cell count, gene number, clusters, and sparsity.}
        \label{fig:combined}
    \end{minipage}
\end{figure*}

\methodname~comprises a curated collection of 13 single-cell RNA-seq datasets from human and mouse, spanning 9 distinct tissue types. 
Comprehensive dataset characteristics, covering species, tissue origin, cell counts, gene dimensionality, annotated clusters, sparsity, and sequencing protocols, are summarized in Fig.2,~\ref{fig:combined} and Tab.~\ref{tab:selected_dataset}.
Specifically, cell counts range from 611 to 22,592, including three large-scale datasets with more than 10,000 cells. 
Gene dimensionality varies from 4,999 to 61,759, with six high-dimensional datasets exceeding 60,000 genes. 
The number of annotated cell types per dataset ranges from 2 to 39, with four datasets containing at least 20 clusters, reflecting substantial cellular heterogeneity. 
Sparsity is generally high, with most datasets (12/13) exceeding 80\% and overall values ranging from 73.02\% to 95.42\%.
The datasets were generated using diverse experimental protocols, including Smart-seq2, 10X Genomics, and CEL-seq2, capturing a wide spectrum of technical platforms and biological contexts. 
Taken together, these characteristics underscore the diversity and comprehensiveness of~\methodname, establishing it as a standardized, AI-ready resource for cell clustering, cell type annotation, marker gene identification, cross-dataset benchmarking, and other downstream single-cell analysis tasks.

\subsection{Dataset Format}
All scRNA-seq datasets are distributed in the \texttt{.h5ad} format, ensuring compatibility with widely used single-cell analysis frameworks such as Scanpy while supporting efficient data storage and manipulation.
Datasets can be readily accessed using \texttt{scanpy.read\_h5ad("path/to/file.h5ad")}, which returns an AnnData object.
Within this object, multiple layers of information are organized in a standardized manner, and the unified structure comprises:

% Each dataset contains the following key components:
\begin{itemize}
    \item \textbf{Gene names:} 
    
    stored in \texttt{data.var["feature\_name"]}, representing standardized gene identifiers.
    \item \textbf{Cell type annotations:} 
    
    recorded in \texttt{data.obs["cell\_type"]}, providing ground-truth labels for each cell.
    \item \textbf{Expression matrix:} 
    
    stored in \texttt{adata.X} (cells-by-genes), which can be converted to a Pandas DataFrame if needed.
\end{itemize}

The standardized data structure ensures consistent retrieval of both expression profiles and cell type annotations, thereby supporting reproducible benchmarking across heterogeneous datasets. 
The dataset can be directly inspected via \texttt{print(adata)}, which displays the available keys across major components, including \texttt{.obs}, \texttt{.var}, \texttt{.uns}, and \texttt{.obsm}. 
These attributes respectively encode cell-level and gene-level metadata, raw and normalized expression matrices, as well as derived features such as PCA or UMAP embeddings. 
By integrating multiple layers of information within a unified framework, this representation offers transparent, versatile, and reproducible access for downstream analyses.

\subsection{Data Preprocessing}
In this study, we implemented a rigorous and standardized preprocessing pipeline to ensure consistency and comparability across single-cell RNA sequencing datasets. Raw data were loaded from \texttt{.h5ad} files into AnnData objects, which contain gene expression matrices along with corresponding metadata and cell annotations. The preprocessing workflow systematically examined whether datasets had undergone normalization, log-transformation (log1p), and scaling. For datasets lacking normalization, library size normalization was applied to mitigate variability associated with sequencing depth, followed by the computation of cell-specific size factors to standardize expression levels across cells. Untransformed data were subjected to log1p transformation to reduce skewness in expression distributions. Finally, z-score scaling was applied to center and scale gene expression values, yielding features with uniform variance. 
This standardized pipeline ensures data quality and stability, while providing a standardized input foundation that supports reproducible and comparable downstream single-cell analyses.

Overall, \methodname~provides a high-quality, AI-ready resource that integrates diverse large-scale datasets from multiple experimental platforms. Its standardized preprocessing and consistent annotation make it well-suited for AI model development, fair method evaluation, and systematic single-cell analyses, thereby establishing a robust foundation for advancing computational single-cell research.

\section{Experiment}

\subsection{Experiment Setup}
\subsubsection{Baseline Methods}
To illustrate the versatility of~\methodname, we assessed its ability to support multiple analytical paradigms by applying three representative clustering algorithms and a biological foundation model designed for classification.
Overall, these methods span a diverse methodological spectrum, from traditional community detection to deep learning and foundation models, thereby underscoring the broad applicability of~\methodname~as a unified resource across different computational paradigms.
The purpose of this evaluation is not to conduct an exhaustive performance comparison among methods, but rather to showcase that the AI-ready datasets in~\methodname~make it possible to apply different modeling approaches in a consistent and reproducible manner.
Specifically, we include the following methods.
\begin{itemize}
    \item \textbf{Leiden.} A graph-based clustering method implemented in \textit{Seurat} (R) that improves upon Louvain to produce stable and well-resolved partitions~\cite{stuart2019comprehensive}. The number of clusters is determined automatically.
    \item \textbf{scMAE.} A masked autoencoder framework for scRNA-seq implemented in Python that reconstructs perturbed expression profiles to learn robust latent cell representations, enabling flexible clustering resolution through adjustable hyperparameters~\cite{fang2024scmae}.  
    \item \textbf{scCDCG.} A deep graph clustering framework implemented in Python that integrates graph construction, self-supervised representation learning, and autoencoder-based feature extraction to capture higher-order structural signals, offering adjustable clustering resolution through hyperparameter tuning~\cite{xu2024sccdcg}.
    \item \textbf{GeneCompass.} A knowledge-informed cross-species foundation model implemented in Python that leverages biological priors to characterize gene regulation and cell-state transitions, supporting fine-tuning for downstream classification tasks~\cite{yang2024genecompass}.
\end{itemize}

% Overall, these methods span a diverse methodological spectrum, from traditional community detection to deep learning and foundation models, thereby underscoring the broad applicability of~\methodname~as a unified resource across diverse computational paradigms.

\subsubsection{Implementation Details}
All methods were run with the parameter settings recommended in their original publications. 
When such settings were not specified, minimal adjustments were applied to ensure stable execution. 
Each method–dataset pair was evaluated over five independent runs, and the mean result was reported.

\subsubsection{Evaluation Metics}
% To evaluate the performance of both clustering and classification methods, we employ four widely used metrics: Accuracy (ACC), Normalized Mutual Information (NMI), Adjusted Rand Index (ARI), and the F1 score~\cite{xia2014robust}. 
% For each metric, the larger value implies a better result.
% ACC measures the proportion of correctly assigned cells, either by comparing predicted clusters to ground-truth labels in unsupervised clustering or by direct label prediction in supervised classification. 
% NMI quantifies the agreement between predicted and true labels by measuring shared information, while ARI assesses the similarity of assignments while adjusting for chance. 
% The F1 score, as the harmonic mean of precision and recall, captures the balance between false positives and false negatives in both clustering and classification contexts. 
To comprehensively evaluate the performance of clustering and classification methods, we adopted two distinct sets of evaluation metrics. 
For clustering assessment, three widely adopted indices were considered: Accuracy (ACC), Normalized Mutual Information (NMI), and Adjusted Rand Index (ARI)~\cite{xia2014robust}. 
Specifically, ACC quantifies the proportion of correctly assigned cells by aligning predicted clusters with ground-truth labels, 
NMI measures the amount of shared information between predicted and true partitions, 
ARI evaluates assignment similarity while correcting for random chance. 
For classification assessment, three standard metrics were adopted: Accuracy (ACC), Precision (PRE), and Recall (REC). 
Here, ACC measures the overall correctness of label prediction, 
PRE reflects the fraction of predicted positives that are true positives, 
and REC indicates the proportion of true positives successfully recovered.

\begin{table}[!t]
    \centering
    \scriptsize
    % \vspace{-5mm}
    \setlength{\tabcolsep}{1.2pt}
    \renewcommand{\arraystretch}{0.9}
    \begin{tabular}{lccccccc}
    \toprule
    \multirow{3}{*}{\textbf{Dataset}} & \multirow{3}{*}{\textbf{Metrics}} & \multicolumn{3}{c}{\textbf{Clustering}} &  & \multirow{3}{*}{\textbf{Metrics}} & \textbf{Classification} \\ 
    \cmidrule{3-5} \cmidrule{8-8}
    & & \textbf{Leiden} & \textbf{scMAE} & \textbf{scCDCG} &  &  & \textbf{Genecompass} \\
    \midrule
    \multirow{4}{*}{Mauro Pancreas} 
         & ACC & 92.08±{\tiny 0.13} & 95.62±{\tiny 0.16} & 92.65±{\tiny 2.93} & & ACC & 98.35±{\tiny 0.17}
 \\      
         & NMI & 89.96±{\tiny 0.12} & 88.49±{\tiny 0.44} & 86.81±{\tiny 0.98} & & PRE & 97.29±{\tiny 0.27}
 \\
         & ARI & 93.57±{\tiny 0.12} & 92.38±{\tiny 0.35} & 91.37±{\tiny 1.21} & & REC & 98.26±{\tiny 0.13}
 % \\
         % & F1  & 85.42±{\tiny 0.19} &                    &                    & & F1  & 97.74±{\tiny 0.18}
 \\
    \midrule
    \multirow{4}{*}{Sonya Liver} 
         & ACC & 69.84±{\tiny 5.17} & 80.73±{\tiny 1.86} & 75.34±{\tiny 3.67} & & ACC & 98.58±{\tiny 0.14}
 \\
         & NMI & 70.70±{\tiny 0.05} & 85.59±{\tiny 1.44} & 71.34±{\tiny 2.59} & & PRE & 98.39±{\tiny 0.31}
 \\
         & ARI & 54.86±{\tiny 0.53} & 88.92±{\tiny 1.78} & 81.26±{\tiny 2.69} & & REC & 97.73±{\tiny 0.50}
 \\
         % & F1  & 71.98±{\tiny 4.44} &                    &                    & & F1  &  \\
    \midrule
    \multirow{4}{*}{Sapiens Liver} 
         & ACC & 71.19±{\tiny 0.00} & 67.51±{\tiny 2.30} & 73.09±{\tiny 1.40} & & ACC & 87.78±{\tiny 1.21} \\
         & NMI & 70.15±{\tiny 0.00} & 78.25±{\tiny 0.60} & 62.54±{\tiny 3.20} & & PRE & 71.68±{\tiny 1.17}
 \\
         & ARI & 48.05±{\tiny 0.00} & 63.82±{\tiny 4.10} & 41.11±{\tiny 4.40} & & REC & 72.52±{\tiny 3.03}
 \\
         % & F1  & 57.71±{\tiny 0.00} &                    &                    & & F1  &  \\
    \midrule
    \multirow{4}{*}{\parbox{2cm}{Sapiens Ear \\ Crista Ampullaris}} 
         & ACC & 43.37±{\tiny 0.94} & 67.00±{\tiny 0.40} & 85.45±{\tiny 4.30} & & ACC & 94.92±{\tiny 1.04} \\
         & NMI & 73.42±{\tiny 0.63} & 74.31±{\tiny 0.20} & 69.54±{\tiny 2.90} & & PRE & 93.42±{\tiny 3.84}
 \\
         & ARI & 59.38±{\tiny 0.31} & 57.35±{\tiny 0.10} & 66.16±{\tiny 4.80} & & REC & 85.34±{\tiny 2.26}
 % \\
         % & F1  & 66.61±{\tiny 0.03} &                    &                    & & F1  & 88.50±{\tiny 2.46}
 \\
    \midrule
    \multirow{4}{*}{Sapiens Ear Utricle} 
         & ACC & 51.21±{\tiny 0.00} & 73.16±{\tiny 0.60} & 79.58±{\tiny 0.70} & & ACC & 98.06±{\tiny 1.77} \\
         & NMI & 71.20±{\tiny 0.00} & 78.28±{\tiny 1.20} & 66.82±{\tiny 5.40} & & PRE & 98.46±{\tiny 2.15}
 \\
         & ARI & 62.76±{\tiny 0.00} & 64.09±{\tiny 0.70} & 60.16±{\tiny 3.30} & & REC & 94.80±{\tiny 4.18}
 \\
         % & F1  & 56.12±{\tiny 0.00} &                    &                    & & F1  &  \\
    \midrule
    \multirow{4}{*}{Sapiens Lung} 
         & ACC & 48.38±{\tiny 0.11} & 63.24±{\tiny 1.10} & 62.06±{\tiny 1.60} & & ACC & 87.44±{\tiny 1.61} \\                       
         & NMI & 70.16±{\tiny 0.17} & 79.65±{\tiny 0.60} & 66.94±{\tiny 1.90} & & PRE & 77.78±{\tiny 2.91}
 \\
         & ARI & 44.19±{\tiny 0.37} & 58.40±{\tiny 1.60} & 60.15±{\tiny 1.50} & & REC & 77.14±{\tiny 2.41}
 \\
         % & F1  & 50.67±{\tiny 0.10} &                    &                    & & F1  &  \\
    \midrule
    \multirow{4}{*}{Sapiens Testis} 
         & ACC & 62.86±{\tiny 0.00} & 53.71±{\tiny 0.80} & 67.18±{\tiny 3.80} & & ACC & 97.33±{\tiny 0.38}
 \\                       
         & NMI & 44.26±{\tiny 0.00} & 57.09±{\tiny 0.40} & 57.42±{\tiny 3.30} & & PRE & 94.02±{\tiny 2.08}
 \\
         & ARI & 51.52±{\tiny 0.00} & 43.27±{\tiny 0.50} & 55.38±{\tiny 7.20} & & REC & 88.03±{\tiny 0.84}
 \\
         % & F1  & 34.02±{\tiny 0.00} &                    &                    & & F1  &  \\
    \midrule
    \multirow{4}{*}{Sapiens Trachea} 
         & ACC & 48.49±{\tiny 0.95} & 65.78±{\tiny 3.70} & 52.46±{\tiny 2.90} & & ACC & 98.21±{\tiny 0.09}
 \\                       
         & NMI & 63.81±{\tiny 1.62} & 77.12±{\tiny 1.50} & 63.25±{\tiny 1.00} & & PRE & 91.17±{\tiny 1.32}
 \\
         & ARI & 41.80±{\tiny 2.34} & 53.69±{\tiny 4.80} & 42.92±{\tiny 2.40} & & REC & 91.06±{\tiny 0.65}
 \\
         % & F1  & 55.10±{\tiny 0.00} &                    &                    & & F1  &  \\
    \midrule
    \multirow{4}{*}{Muris Limb Muscle} 
         & ACC & 96.72±{\tiny 1.12} & 66.13±{\tiny 3.40} & 94.50±{\tiny 7.10} & & ACC & 96.63±{\tiny 0.76} \\                       
         & NMI & 0.24±{\tiny 0.11} & 59.44±{\tiny 3.80} & 56.54±{\tiny 7.60} & & PRE & 94.66±{\tiny 1.26}
 \\
         & ARI & 8.54±{\tiny 0.66} & 51.54±{\tiny 3.30} & 53.37±{\tiny 8.50} & & REC & 94.73±{\tiny 1.35}
 \\
         % & F1  & 63.95±{\tiny 0.11} &                    &                    & & F1  &  \\
    \midrule
    \multirow{4}{*}{Muris Brain} 
         & ACC & 40.84±{\tiny 0.01} & 71.37±{\tiny 0.00} & 95.55±{\tiny 1.10} & & ACC & 100.00±{\tiny 0.00} \\                       
         & NMI & 26.52±{\tiny 0.01} & 1.33±{\tiny 0.00} & 22.48±{\tiny 8.30} & & PRE & 100.00±{\tiny 0.00}
 \\
         & ARI & 1.46±{\tiny 0.01} & 2.22±{\tiny 0.00} & 35.56±{\tiny 7.80} & & REC & 100.00±{\tiny 0.00}
 \\
         % & F1  & 49.46±{\tiny 0.00} &                    &                    & & F1  &  \\
    \midrule
    \multirow{4}{*}{Muris Kidney} 
         & ACC & 38.16±{\tiny 1.22} & 55.52±{\tiny 3.40} & 80.65±{\tiny 1.60} & & ACC & 93.85±{\tiny 3.83} \\                       
         & NMI & 21.38±{\tiny 0.58} & 54.37±{\tiny 1.90} & 55.82±{\tiny 1.30} & & PRE & 93.96±{\tiny 3.67}
 \\
         & ARI & 18.43±{\tiny 1.44} & 35.79±{\tiny 1.40} & 42.88±{\tiny 2.10} & & REC & 92.61±{\tiny 3.53}
 \\
         % & F1  & 58.03±{\tiny 0.12} &                    &                    & & F1  &  \\
    \midrule
    \multirow{4}{*}{Muris Liver} 
         & ACC & 45.72±{\tiny 0.14} & 53.48±{\tiny 0.40} & 68.13±{\tiny 1.40} & & ACC & 94.76±{\tiny 0.49}
 \\                       
         & NMI & 50.59±{\tiny 0.10} & 65.39±{\tiny 1.00} & 62.06±{\tiny 2.40} & & PRE & 86.47±{\tiny 0.12}
 \\
         & ARI & 38.63±{\tiny 0.09} & 47.55±{\tiny 0.60} & 46.96±{\tiny 3.70} & & REC & 86.58±{\tiny 1.88}
 \\
         % & F1  & 45.12±{\tiny 0.22} &                    &                    & & F1  &  \\
    \midrule
    \multirow{4}{*}{Muris Lung} 
         & ACC & 50.45±{\tiny 2.96} & 51.06±{\tiny 2.20} & 65.68±{\tiny 1.70} & & ACC & 93.62±{\tiny 1.37}
 \\                       
         & NMI & 64.32±{\tiny 0.46} & 64.49±{\tiny 0.90} & 49.53±{\tiny 3.80} & & PRE & 85.46±{\tiny 3.91}
 \\
         & ARI & 31.22±{\tiny 5.55} & 35.69±{\tiny 2.40} & 26.46±{\tiny 3.80} & & REC & 83.61±{\tiny 2.00}
\\
         % & F1  & 27.99±{\tiny 1.66} &                    &                    & & F1  &  \\
    \bottomrule
    \end{tabular}
    \caption{Performance comparison of clustering and classification methods across datasets.}
    \vspace{-5mm}
    \label{tab:results}
\end{table}

\subsection{Performence}
Table~\ref{tab:results} presents the comparative performance of representative clustering and classification approaches evaluated on the 13 standardized datasets in~\methodname. 

For clustering, Leiden demonstrates robust and consistent performance on relatively small or less complex datasets, e.g., \textit{Mauro Pancreas}. 
However, its accuracy declines markedly when applied to datasets with higher dimensionality or greater cellular diversity, such as \textit{Sapiens Lung} and \textit{Muris Kidney}. 
In contrast, deep learning–based methods, scMAE and scCDCG, exhibit stronger adaptability under these challenging conditions, achieving superior ARI and ACC scores. 
This suggests that models based on representation learning are better suited to capture subtle cellular heterogeneity and complex non-linear structures in single-cell data.
For classification, GeneCompass attains consistently high accuracy across all datasets, often exceeding 95\% and reaching 100\% on \textit{Muris Brain}. 
The high precision and recall further confirm the robustness of GeneCompass, highlighting the effectiveness of foundation models in transferring knowledge from well-annotated references to diverse tissues and species.

Overall, these results highlight the unique value of~\methodname~as a comprehensive and standardized single-cell resource. This unified resource ensures that methodological advances can be assessed systematically and applied broadly, maximizing the impact and comparability of single-cell analyses.

\subsection{Case Study}
\begin{figure}[!t]
\centering
\subfloat[Visualization in 2D projection]{
\includegraphics[height=4cm,width=0.22\textwidth,keepaspectratio]{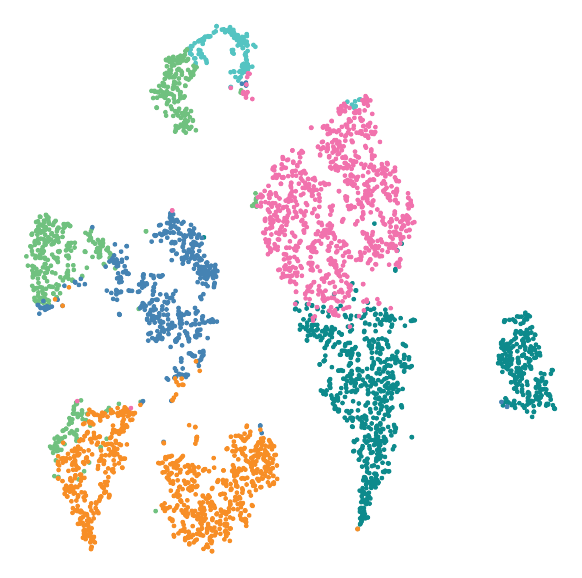}
\label{subfig:sccdcg_vis}
}
\subfloat[Heatmap of cell similarities]{
\includegraphics[height=4cm,width=0.25\textwidth,keepaspectratio]{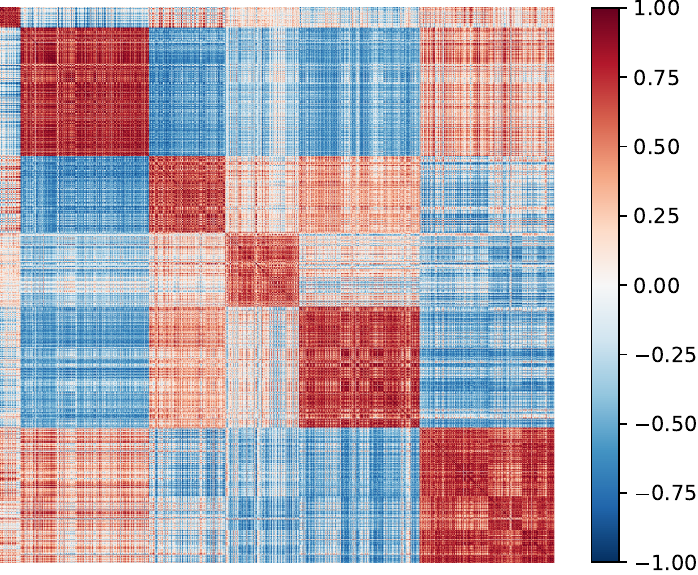}
\label{subfig:sccdcg_heatmap}
}
\\
\vspace{-3mm}
\subfloat[Top 3 DEGs per prediction cluster]{
\includegraphics[width=0.4\textwidth,keepaspectratio]{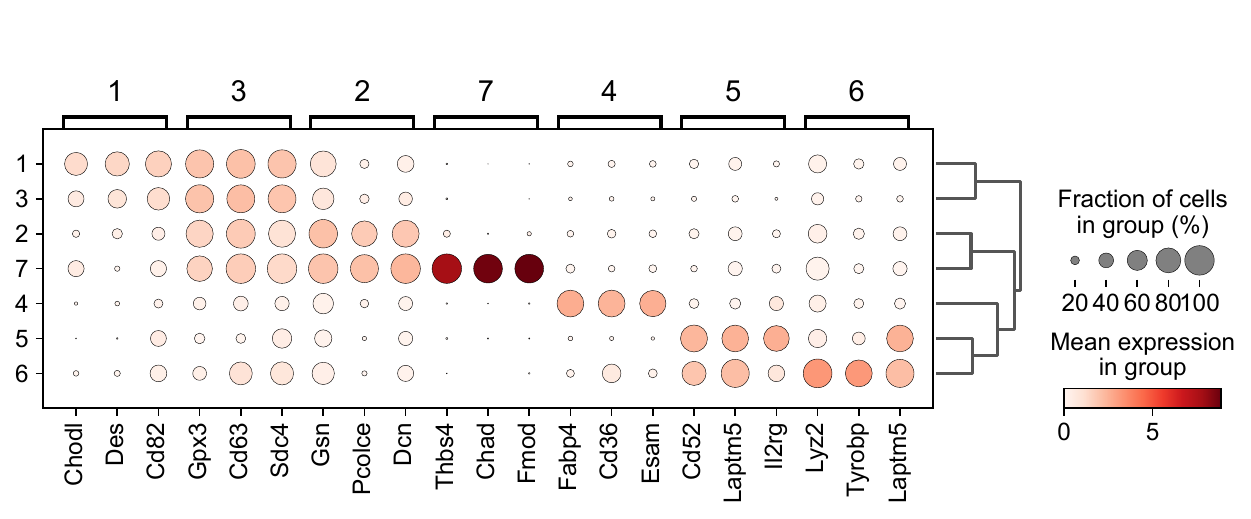}
\label{subfig:sccdcg_dot}
}
\\
% \vspace{-3mm}
\subfloat[Cell type annotation]{
\includegraphics[height=4cm,width=0.45\textwidth,keepaspectratio]{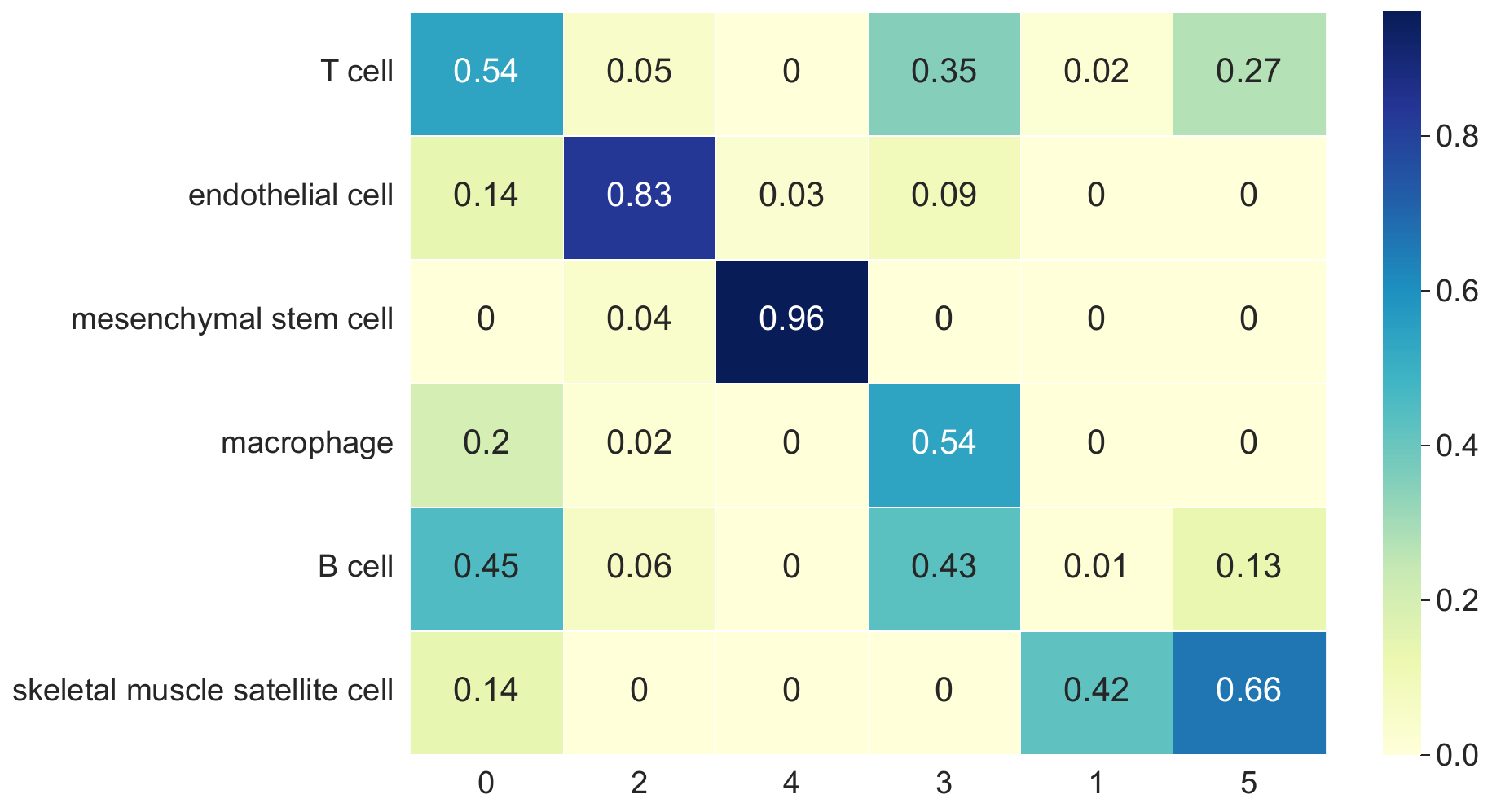}
\label{subfig:sccdcg_cta}
}
\\
\subfloat[Annotation Comparison based on Sankey diagrams]{
\includegraphics[width=0.45\textwidth,keepaspectratio]{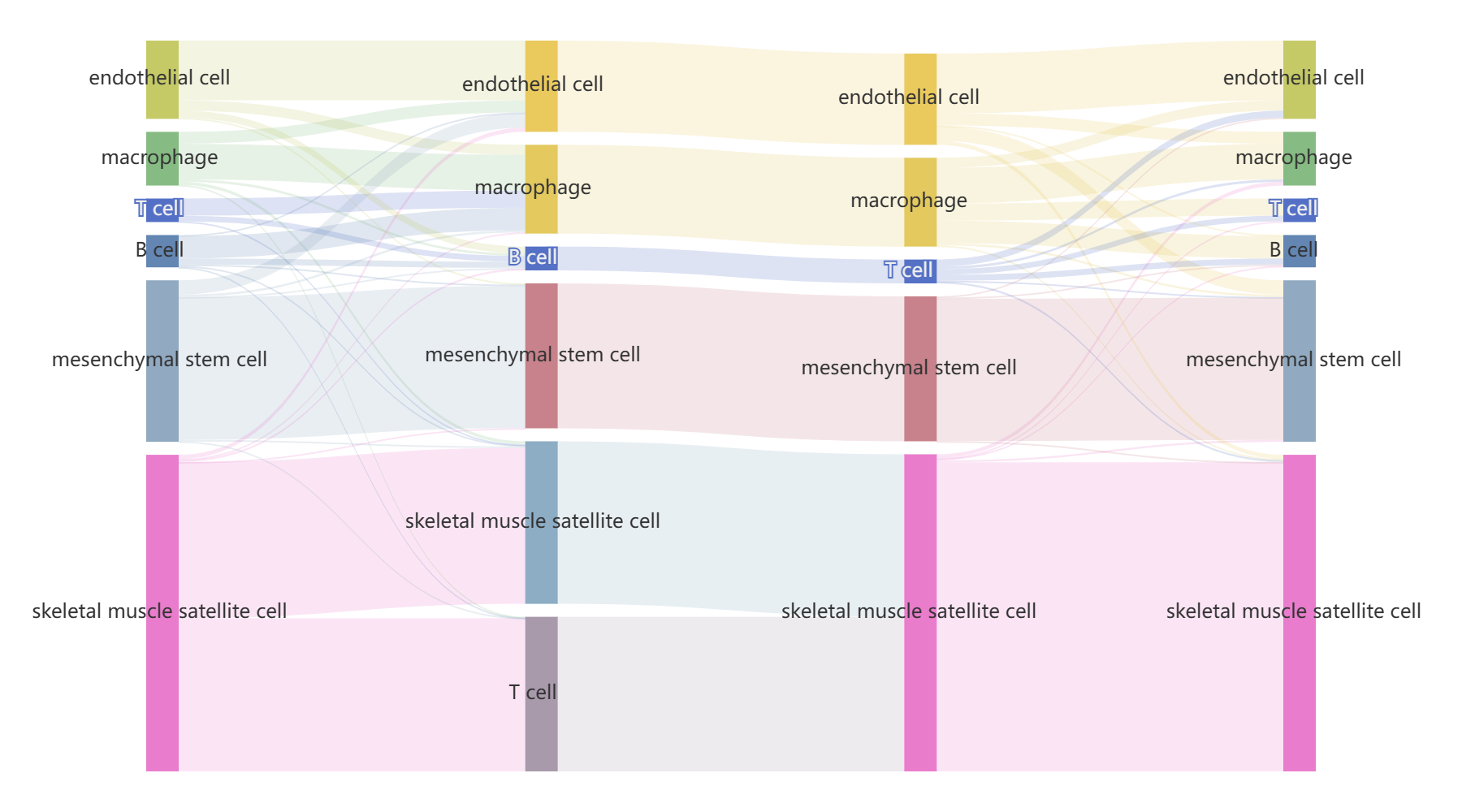}
\label{subfig:sccdcg_sankey}
}
\caption{Case study of scCDCG on \textit{Muris Limb Muscle}, demonstrating integrated representation learning, two-dimensional visualization, and biologically guided cell-type annotation. (e) presents four columns from left to right: Gold-standard labels, results of the Best-mapping annotation, results of the Marker-overlap annotation, and the Gold-standard labels.}
\vspace{-5mm}
\label{fig:casestudy}
\end{figure}

To further illustrate the versatility and practical utility of~\methodname, we present several representative case studies that demonstrate its support for diverse analytical tasks. 
Using the \textit{Muris Limb Muscle} dataset as a primary example, we applied the scCDCG model to learn cell representations, revealing clear separation of cell populations in both cosine similarity heatmaps and two-dimensional projections (Fig.~\ref{fig:casestudy}a-b). 
By integrating prior biological knowledge, we identified highly expressed marker genes informative for cell-type annotation. 
% The resulting cell-type assignments exhibit strong concordance with known labels, as visualized through heatmaps, dot plots of top differentially expressed genes, and Sankey diagrams (Fig.~\ref{fig:casestudy}c-e).
As shown in Fig.~\ref{fig:casestudy}c–e, the resulting cell-type assignments exhibit strong concordance with known labels, demonstrated through heatmaps, dot plots of top differentially expressed genes, and Sankey diagrams.

\subsubsection{Cell Representations and Cluster Visualization}
Accurate clustering depends on the ability to learn discriminative cell representations.
To evaluate this, we first measured the similarity structure of the learned embeddings and visualized it as a heatmap, which highlights strong intra-cluster coherence.
We then employed t-SNE to project the embeddings into a two-dimensional space, offering an intuitive representation of the underlying feature distribution and cluster organization.
As illustrated in Fig.~\ref{subfig:sccdcg_heatmap} and \ref{subfig:sccdcg_vis}, clusters are clearly separated, and cells within each cluster display strong internal consistency.
These results demonstrate that scCDCG effectively preserves transcriptional heterogeneity across cells and yields highly discriminative representations for downstream analysis.

\subsubsection{Marker Gene Identification}
To examine whether the datasets provided by~\methodname~support downstream biological interpretation, we performed differential expression gene (DEG) analysis for each cluster using the \texttt{rank\_genes\_groups} function in Scanpy with default parameters.
For each cluster, the top 100 genes with pronounced cluster-specific expression were selected as candidate marker genes for subsequent cell-type annotation.
As shown in Fig.~\ref{subfig:sccdcg_dot}, the top three representative marker genes of each cluster clearly delineate the expression signatures of distinct cell populations.
For example, Chodl, Des, and Cd82 were identified as marker genes for cluster 1, whereas Gpx3, Cd63, and Sdc4 were identified for cluster 3.

\subsubsection{Cell Type Annotation}
Building upon the identified marker genes, we next evaluated the biological interpretability of the predicted clusters through a marker-overlap annotation strategy. 
Specifically, we first performed differential expression gene (DEG) analysis on the reference clusters provided in the~\methodname~dataset, and retained the top 100 cluster-specific genes to construct a \emph{gold-standard} marker set. 
For each cluster predicted by scCDCG, we quantified its concordance with reference clusters by computing an overlap score as $\text{overlap}(p,g) = \frac{|\text{DEG}_p \cap \text{DEG}_g|}{100},$
where \(p\) and \(g\) denote the predicted and reference clusters, respectively. 
Each predicted cluster was then assigned to the reference cluster with which it shared the highest overlap score, thereby determining its putative cell-type identity. 

As illustrated in Fig.~\ref{subfig:sccdcg_cta}, this approach successfully annotated scCDCG-predicted clusters 2 and 4 as `endothelial cell' and `mesenchymal stem cell', respectively, highlighting the biological coherence of the learned representations and the reliability of~\methodname~for cell-type annotation tasks.

\subsubsection{Annotation Comparison}
Cell type annotation is a central step in scRNA-seq data analysis, and different strategies may yield varying levels of biological interpretability. 
In addition to the \emph{marker-overlap annotation} strategy introduced above, we implemented an alternative approach, termed \emph{best-mapping annotation}. 
This method applies the Hungarian algorithm to establish an optimal one-to-one correspondence between predicted and reference clusters, thereby achieving rapid label alignment independent of gene expression information. 

To systematically compare these strategies and assess their deviation from the gold-standard annotations, we visualized the results using \emph{Sankey diagrams}. 
The direction and width of the flows effectively capture the mapping patterns and degrees of divergence across clusters. 
As illustrated in Fig.~\ref{subfig:sccdcg_sankey}, this comparative analysis demonstrates that biologically informed strategies such as marker-overlap annotation yield more coherent and interpretable results than purely alignment-based approaches, highlighting the importance of integrating biological knowledge into cell-type annotation.

\subsubsection{Validation Across Datasets}
To further validate the quality and utility of the datasets provided by~\methodname, we conducted additional analyses on the \textit{Sapiens Ear Utricle} and \textit{Muris Limb Muscle} datasets. For each dataset, representative methods including Leiden, scMAE, and scCDCG were systematically applied.

\begin{figure*}[!t]
\centering
\subfloat[Leiden]{
\includegraphics[width=0.3\textwidth,keepaspectratio]{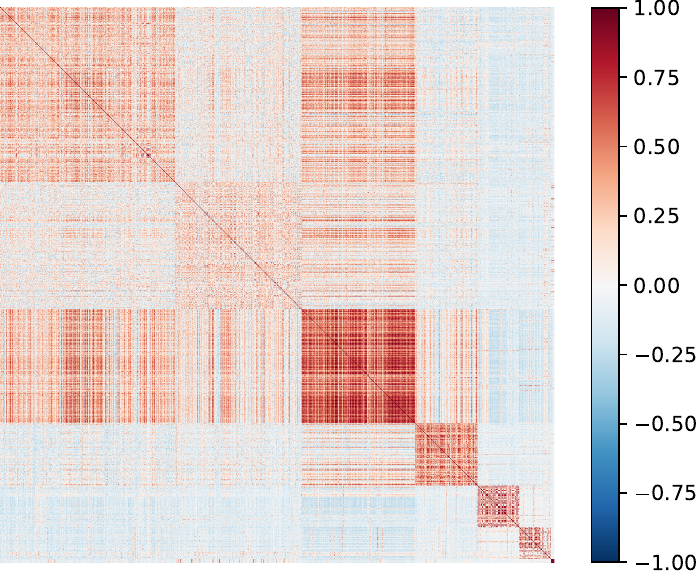}
\label{subfig:}
}
\subfloat[scMAE]{
\includegraphics[width=0.3\textwidth,keepaspectratio]{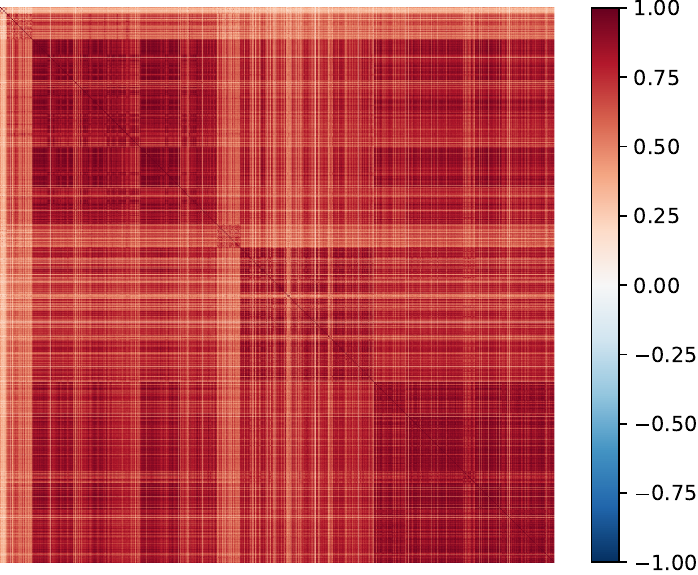}
\label{subfig:}
}
\subfloat[csCDCG]{
\includegraphics[width=0.3\textwidth,keepaspectratio]{image/heat/Tabula_Muris_limb_muscle_filtered_scCDCG.pdf}
\label{subfig:}
}
\\
\subfloat[Leiden]{
\includegraphics[width=0.3\textwidth,keepaspectratio]{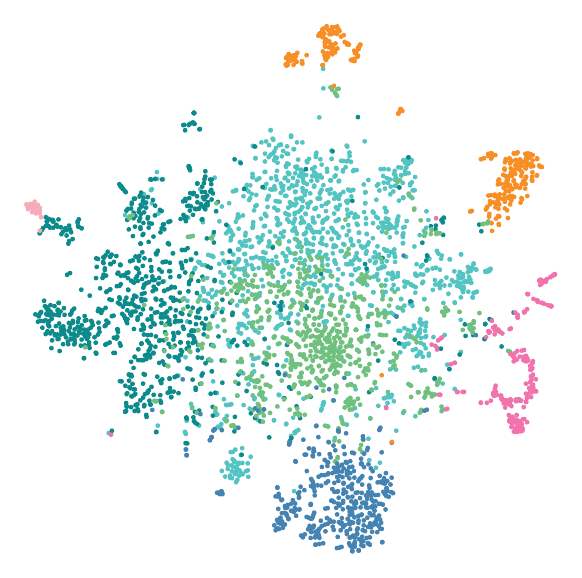}
\label{subfig:}
}
\subfloat[scMAE]{
\includegraphics[width=0.3\textwidth,keepaspectratio]{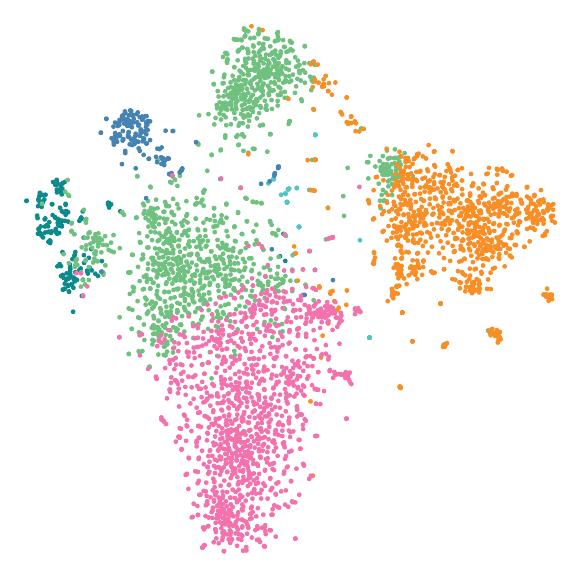}
\label{subfig:}
}
\subfloat[scCDCG]{
\includegraphics[width=0.3\textwidth,keepaspectratio]{image/vis/Tabula_Muris_limb_muscle_filtered_scCDCG.pdf}
\label{subfig:}
}
\\
\subfloat[Leiden]{
\includegraphics[width=0.33\textwidth,keepaspectratio]{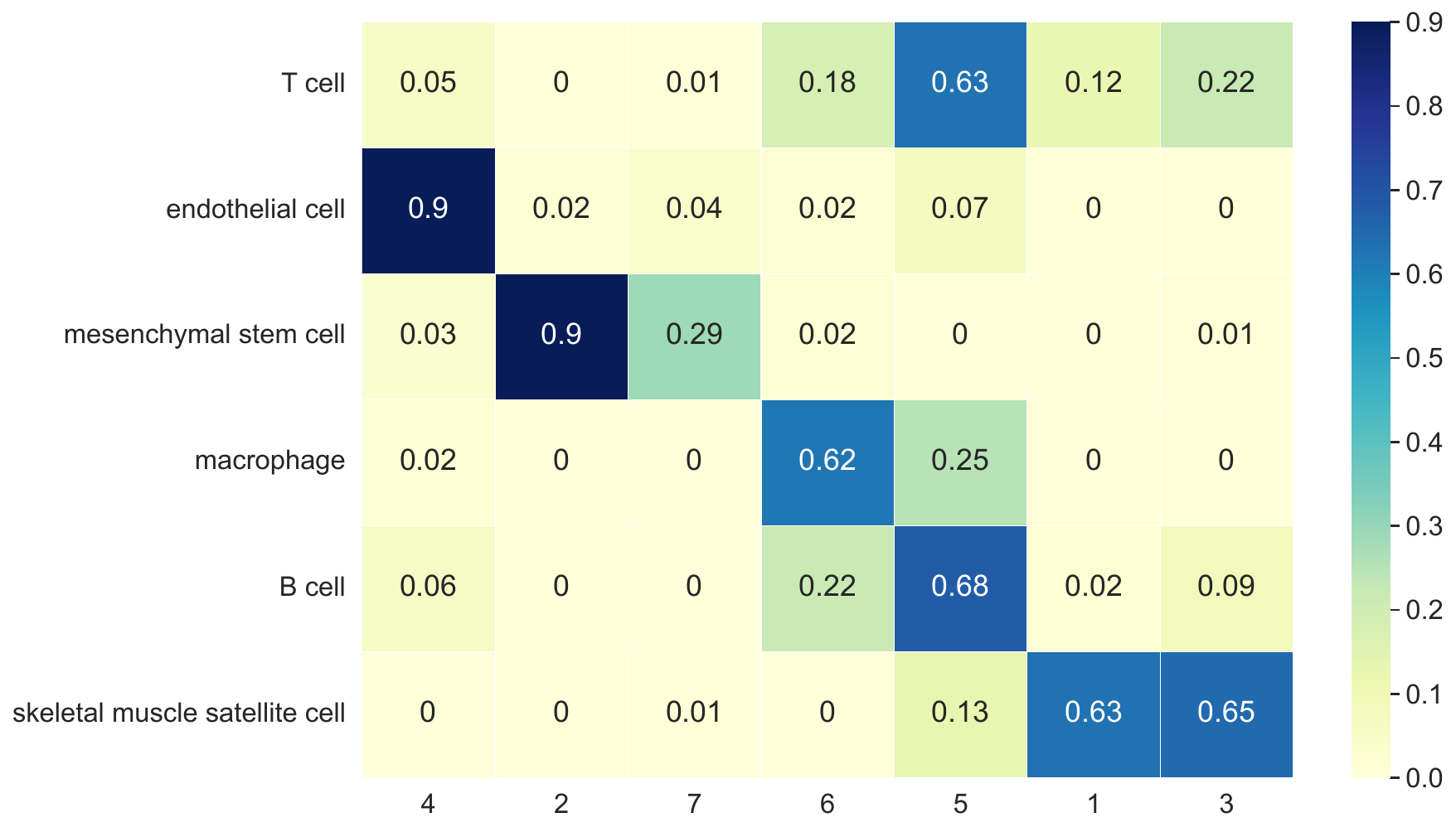}
\label{subfig:}
}
\subfloat[scMAE]{
\includegraphics[width=0.33\textwidth,keepaspectratio]{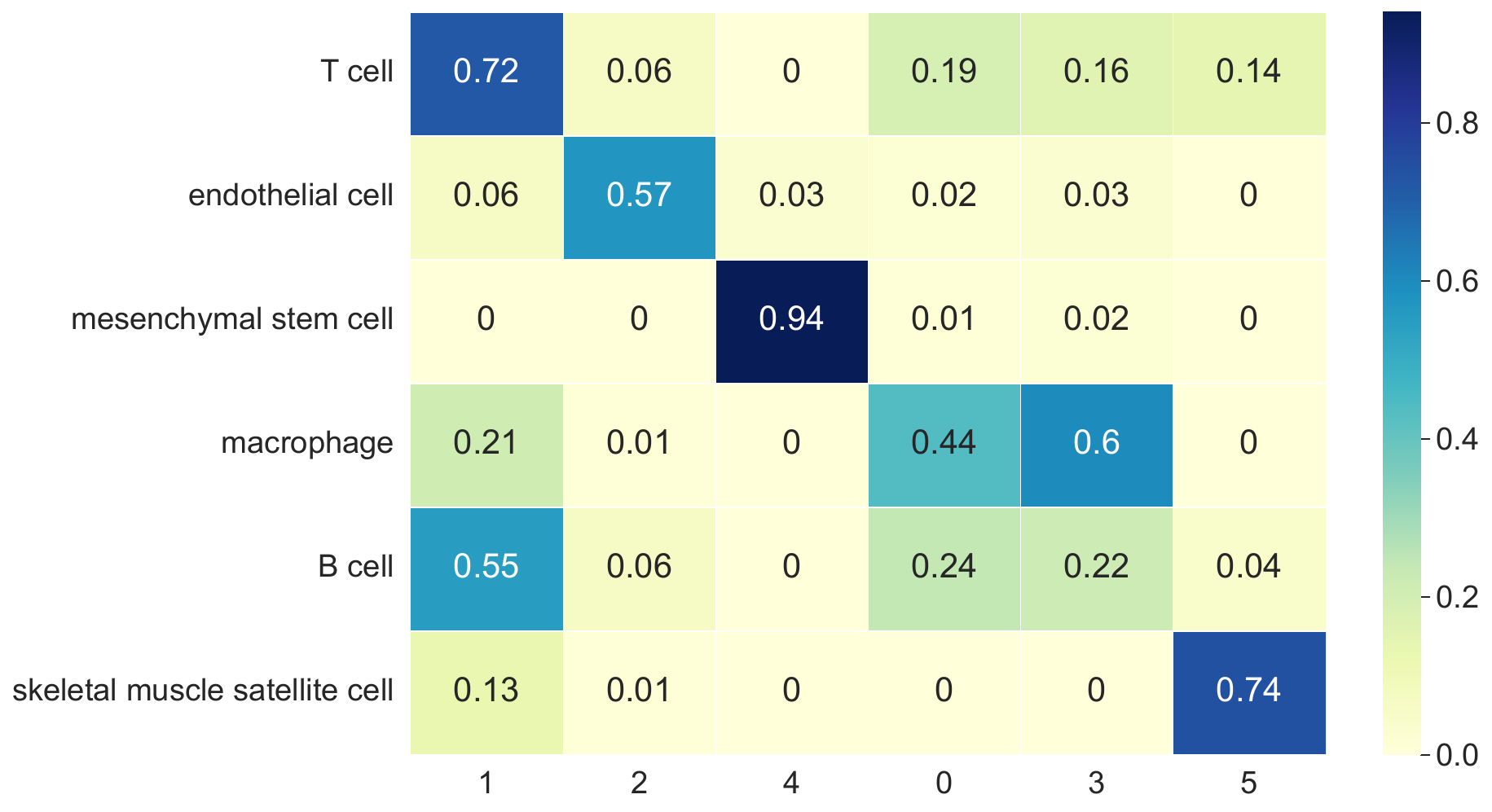}
\label{subfig:}
}
\subfloat[scCDCG]{
\includegraphics[width=0.33\textwidth,keepaspectratio]{image/overlap/Tabula_Muris_limb_muscle_filtered_scCDCG_heatmap.pdf}
\label{subfig:}
}
\\
\subfloat[Leiden]{
\includegraphics[width=0.33\textwidth,keepaspectratio]{image/dot/Tabula_Muris_limb_muscle_filtered_leiden.pdf}
\label{subfig:}
}
\subfloat[scMAE]{
\includegraphics[width=0.33\textwidth,keepaspectratio]{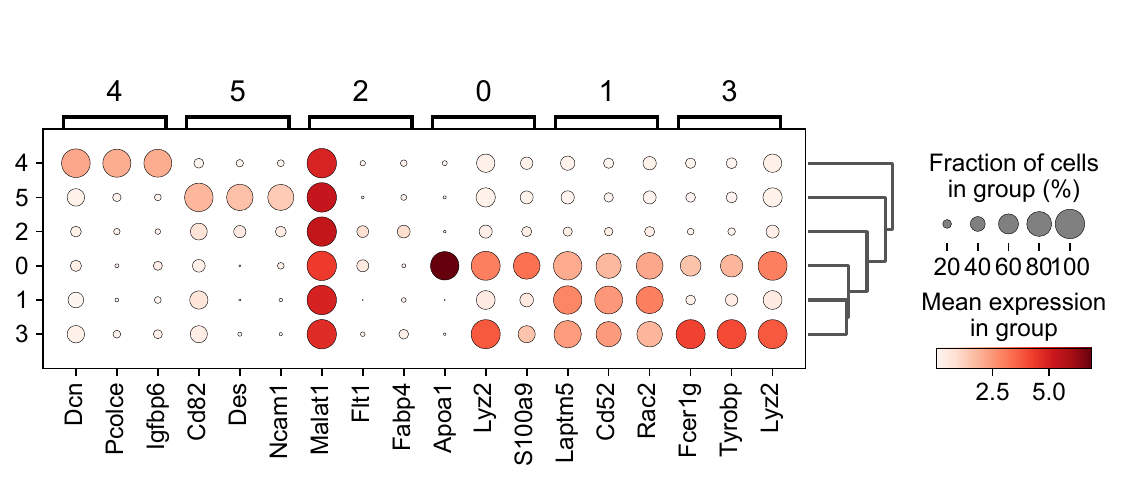}
\label{subfig:}
}
\subfloat[scCDCG]{
\includegraphics[width=0.33\textwidth,keepaspectratio]{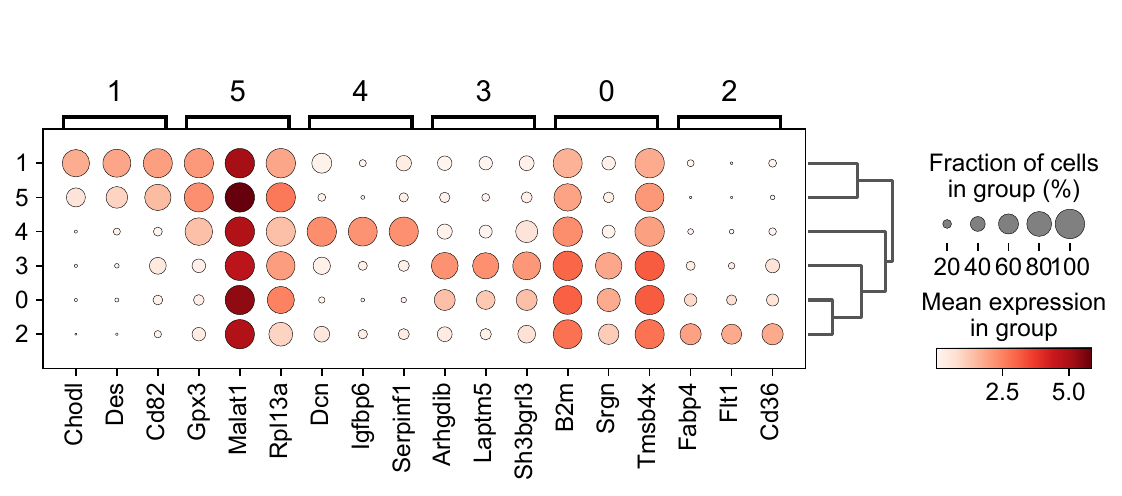}
\label{subfig:}
}
\\
\subfloat[Leiden]{
\includegraphics[width=0.33\textwidth,keepaspectratio]{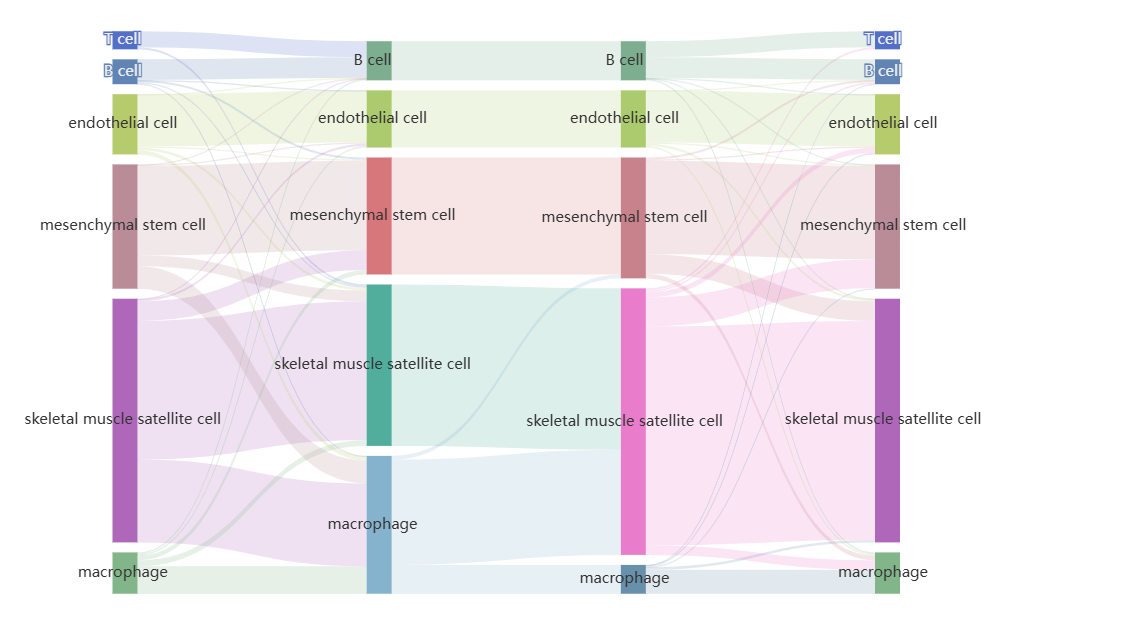}
\label{subfig:}
}
\subfloat[scMAE]{
\includegraphics[width=0.33\textwidth,keepaspectratio]{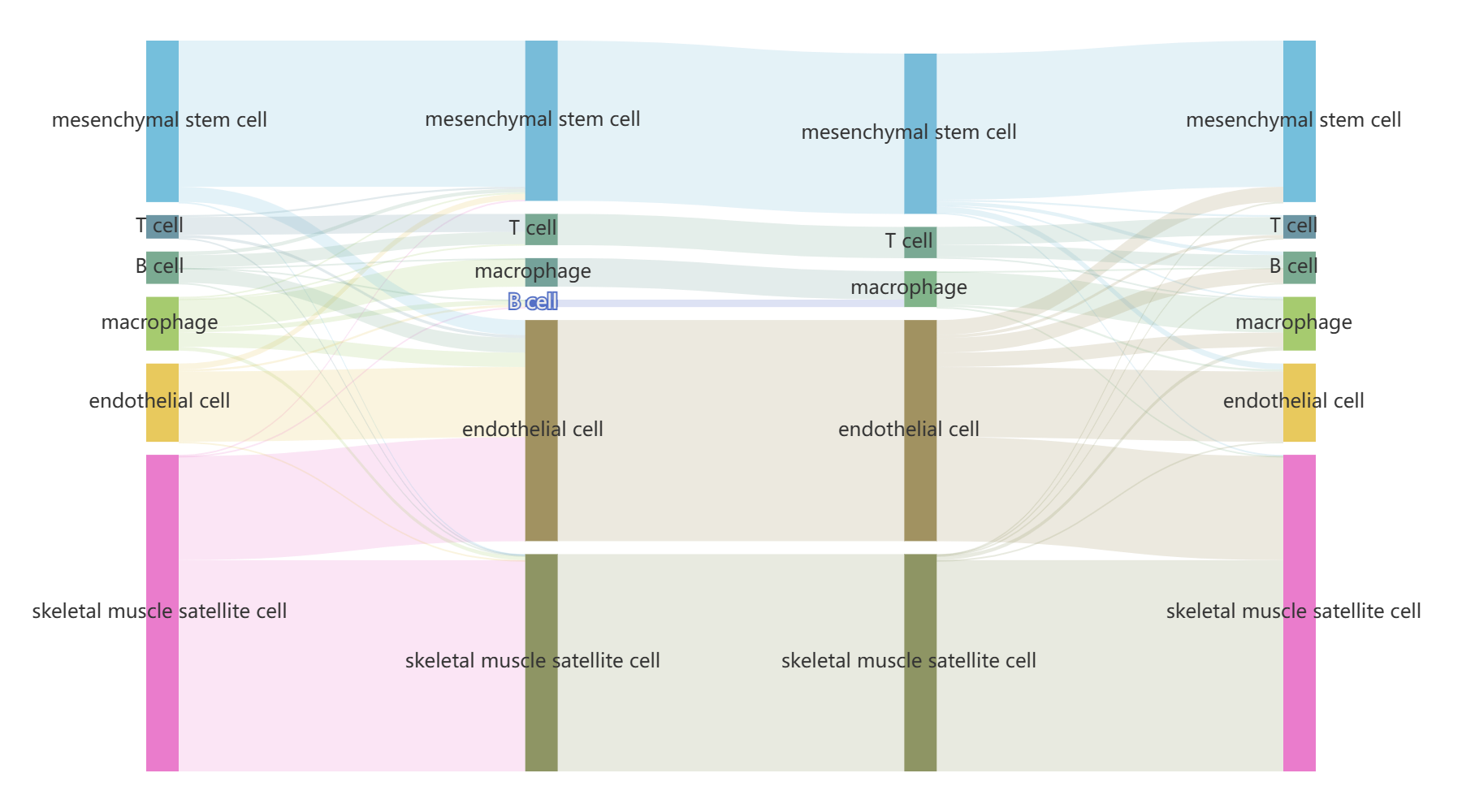}
\label{subfig:}
}
\subfloat[scCDCG]{
\includegraphics[width=0.33\textwidth,keepaspectratio]{image/sankey/sankey_Tabula_Muris_limb_muscle_filtered_scCDCG.png}
\label{subfig:}
}
\caption{Case study on the \textit{Muris Limb Muscle}, summarizing the results of Leiden, scMAE, and scCDCG models, including representation learning, two-dimensional visualization, and marker gene-based cell type annotation. (m)-(o) presents four columns from left to right: Gold-standard labels, results of the Best-mapping annotation, results of the Marker-overlap annotation, and the Gold-standard labels.}
\label{fig:all_muris}
\end{figure*}

\begin{figure*}[!t]
\centering
\subfloat[Leiden]{
\includegraphics[width=0.3\textwidth,keepaspectratio]{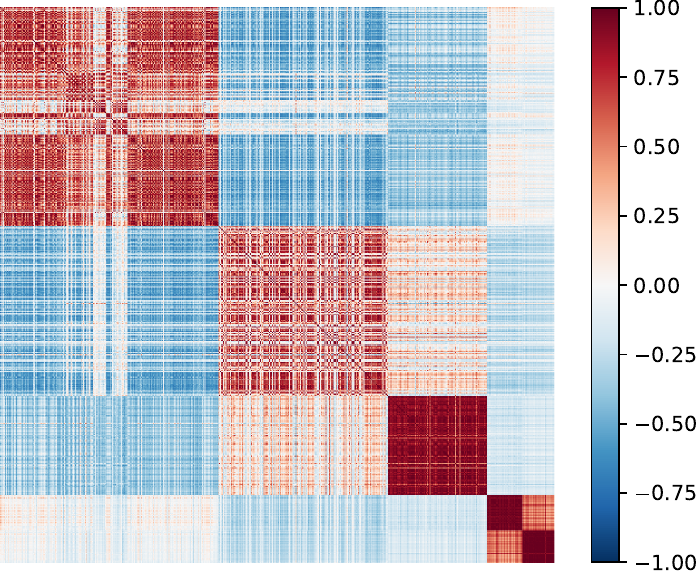}
\label{subfig:}
}
\subfloat[scMAE]{
\includegraphics[width=0.3\textwidth,keepaspectratio]{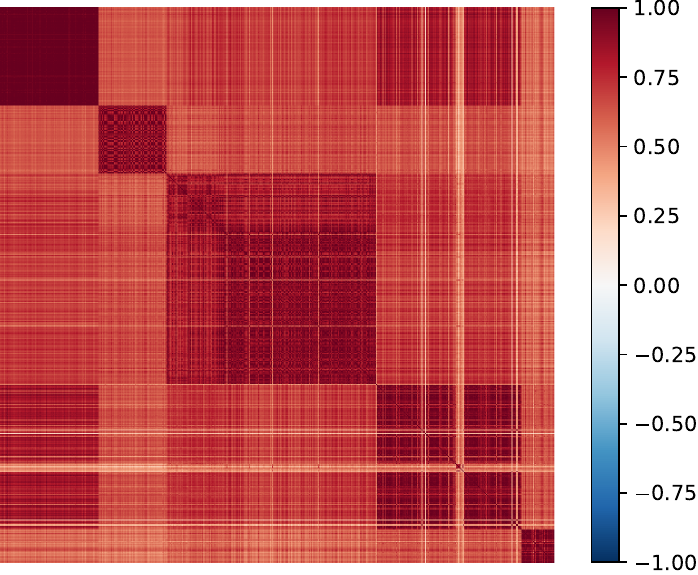}
\label{subfig:}
}
\subfloat[scCDCG]{
\includegraphics[width=0.3\textwidth,keepaspectratio]{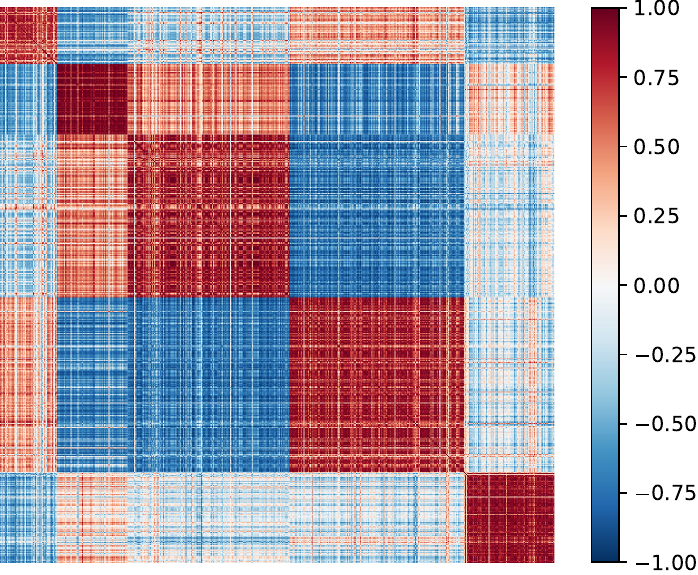}
\label{subfig:}
}
\\
\subfloat[Leiden]{
\includegraphics[width=0.3\textwidth,keepaspectratio]{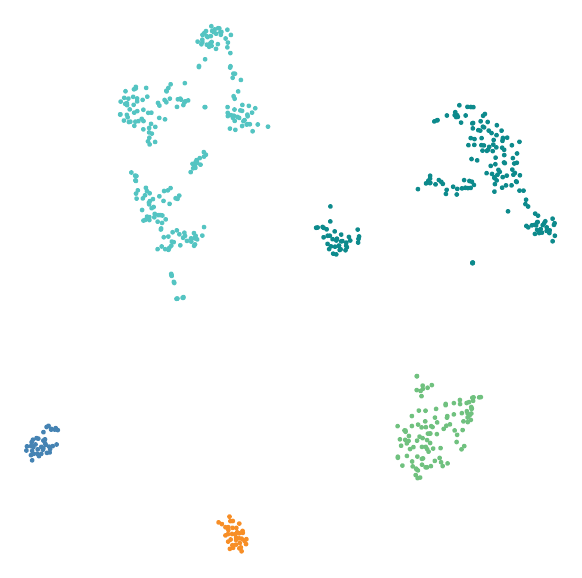}
\label{subfig:}
}
\subfloat[scMAE]{
\includegraphics[width=0.3\textwidth,keepaspectratio]{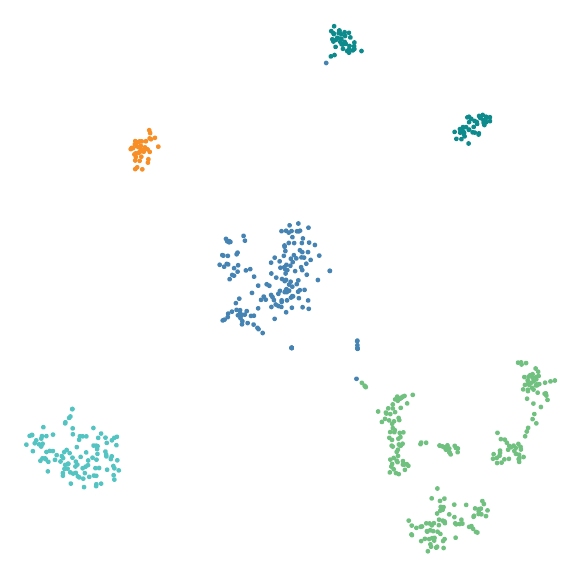}
\label{subfig:}
}
\subfloat[scCDCG]{
\includegraphics[width=0.3\textwidth,keepaspectratio]{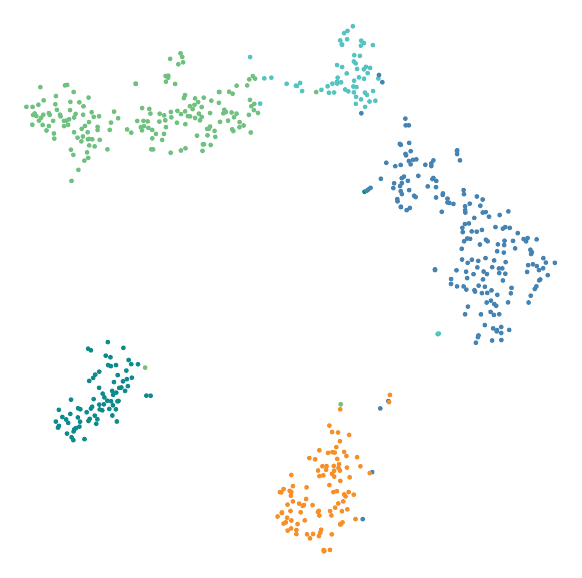}
\label{subfig:}
}
\\
\subfloat[Leiden]{
\includegraphics[width=0.33\textwidth,keepaspectratio]{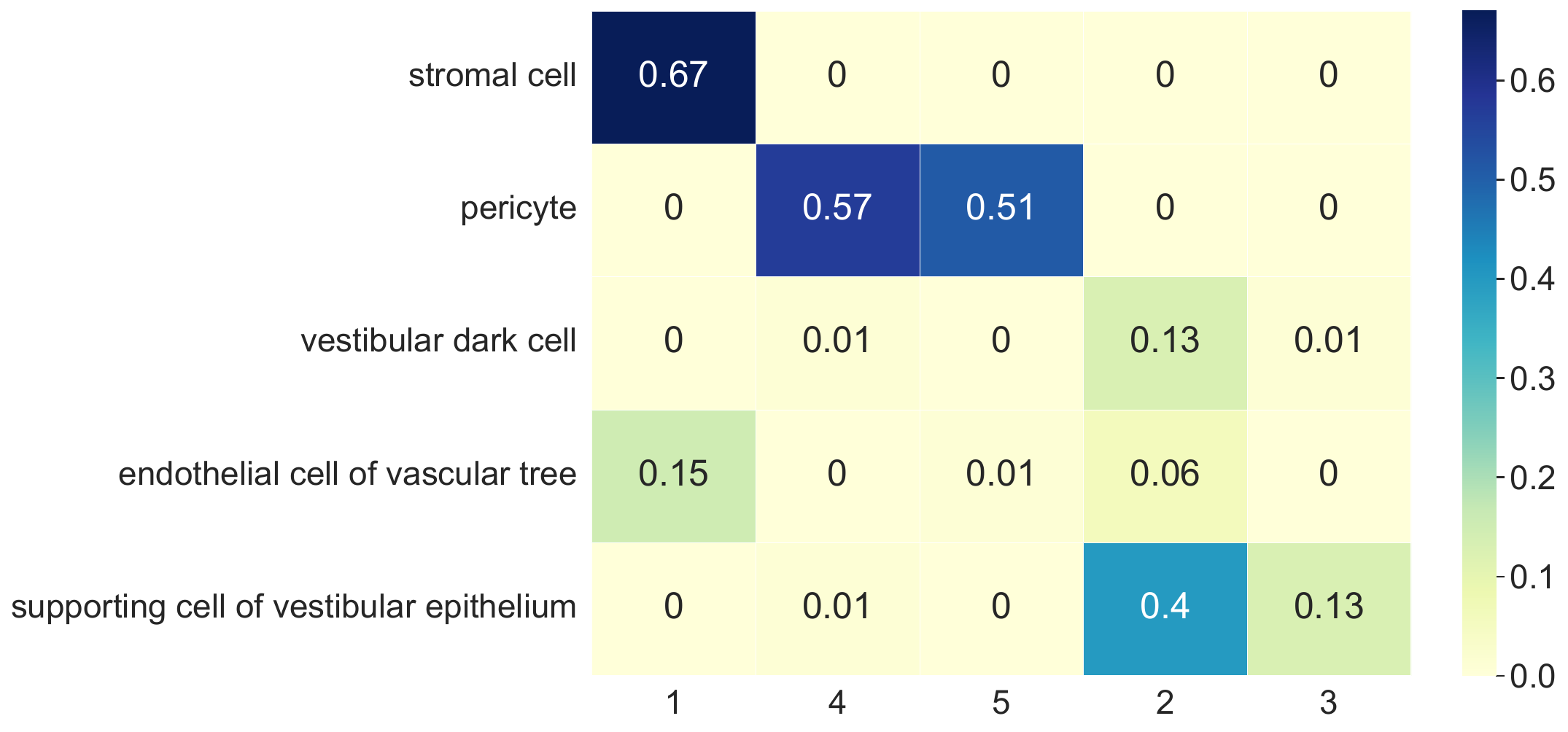}
\label{subfig:}
}
\subfloat[scMAE]{
\includegraphics[width=0.33\textwidth,keepaspectratio]{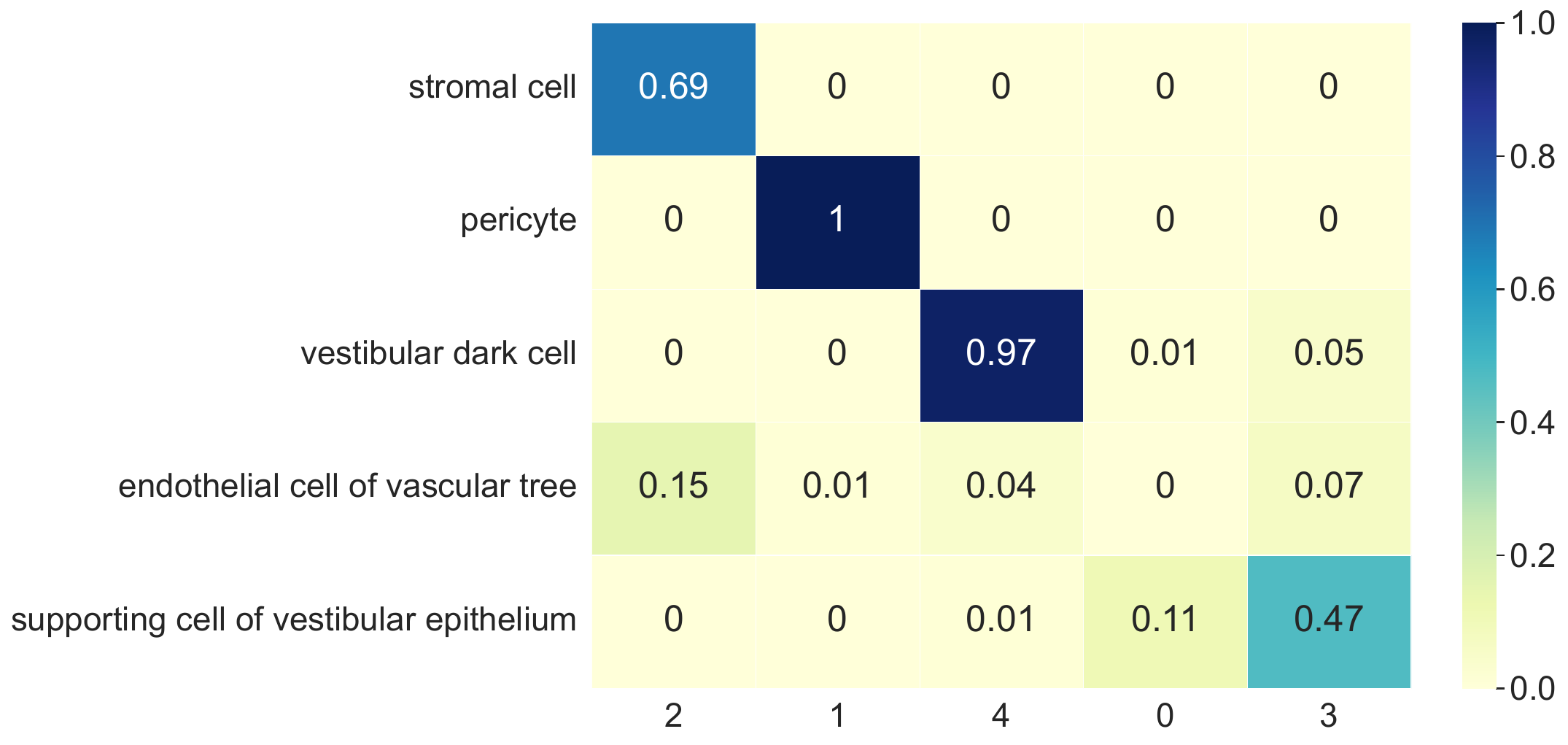}
\label{subfig:}
}
\subfloat[scCDCG]{
\includegraphics[width=0.33\textwidth,keepaspectratio]{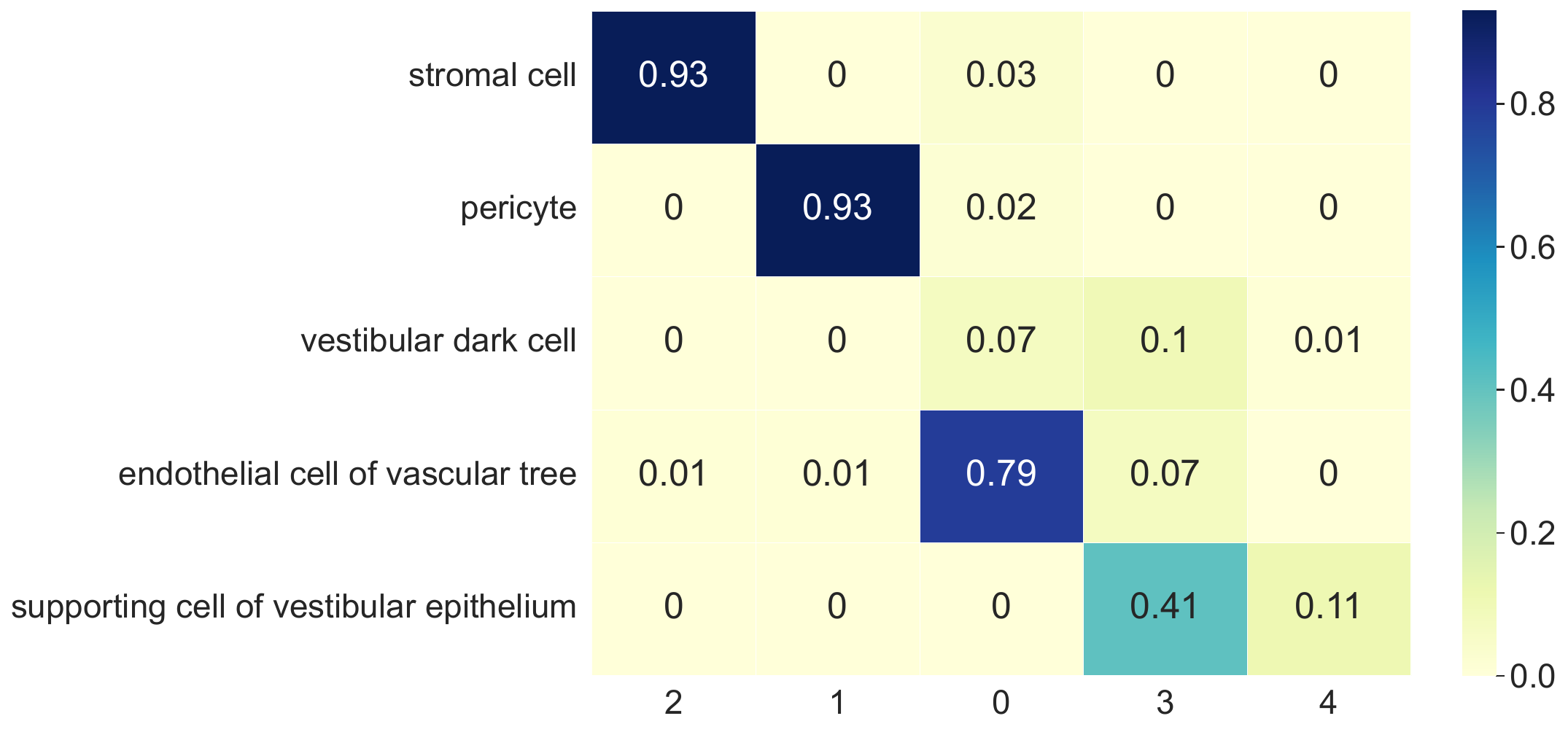}
\label{subfig:}
}
\\
\subfloat[Leiden]{
\includegraphics[width=0.33\textwidth,keepaspectratio]{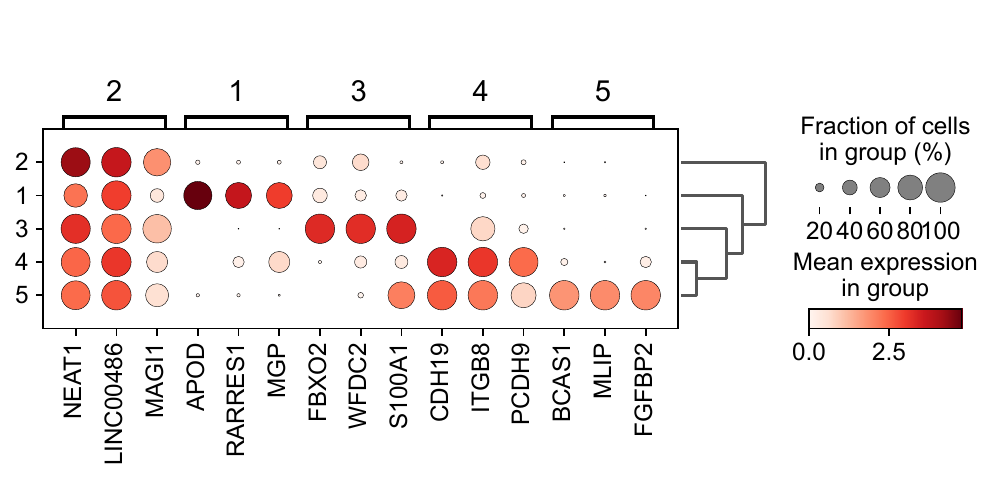}
\label{subfig:}
}
\subfloat[scMAE]{
\includegraphics[width=0.33\textwidth,keepaspectratio]{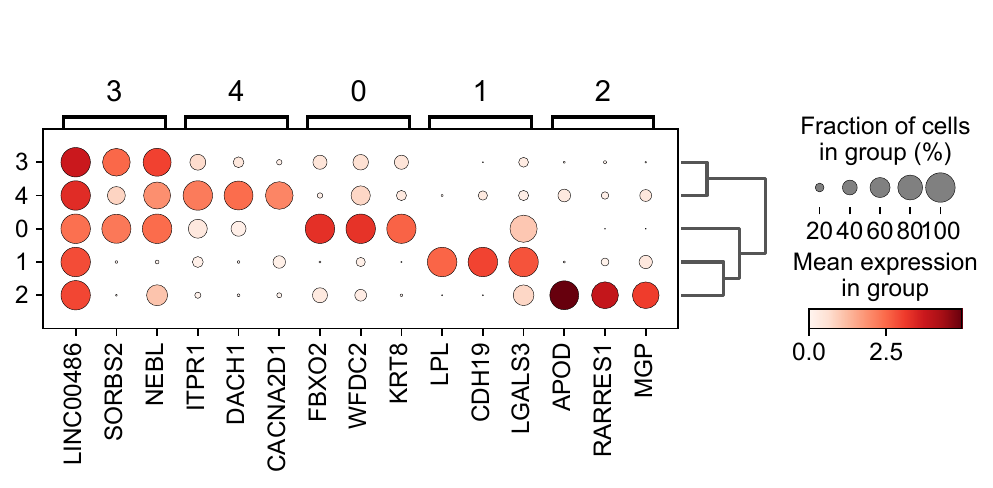}
\label{subfig:}
}
\subfloat[scCDCG]{
\includegraphics[width=0.33\textwidth,keepaspectratio]{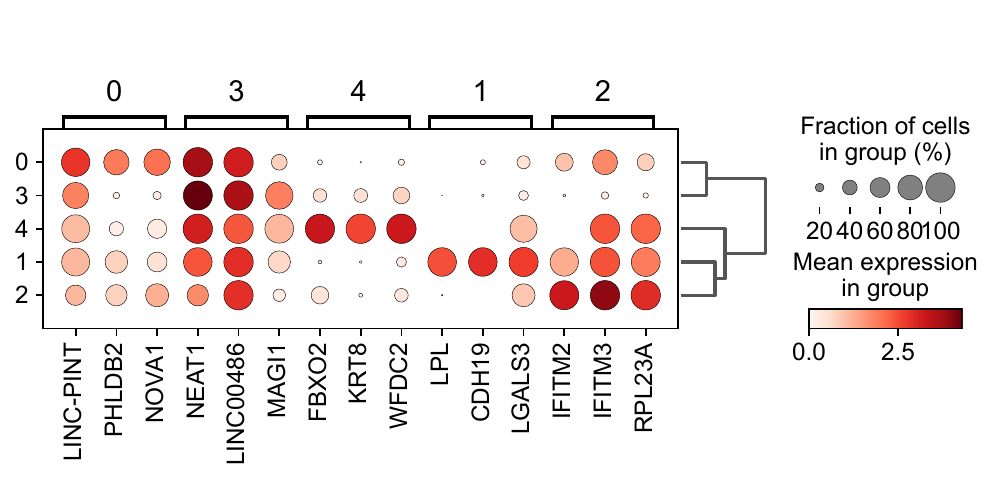}
\label{subfig:}
}
\\
\subfloat[Leiden]{
\includegraphics[width=0.33\textwidth,keepaspectratio]{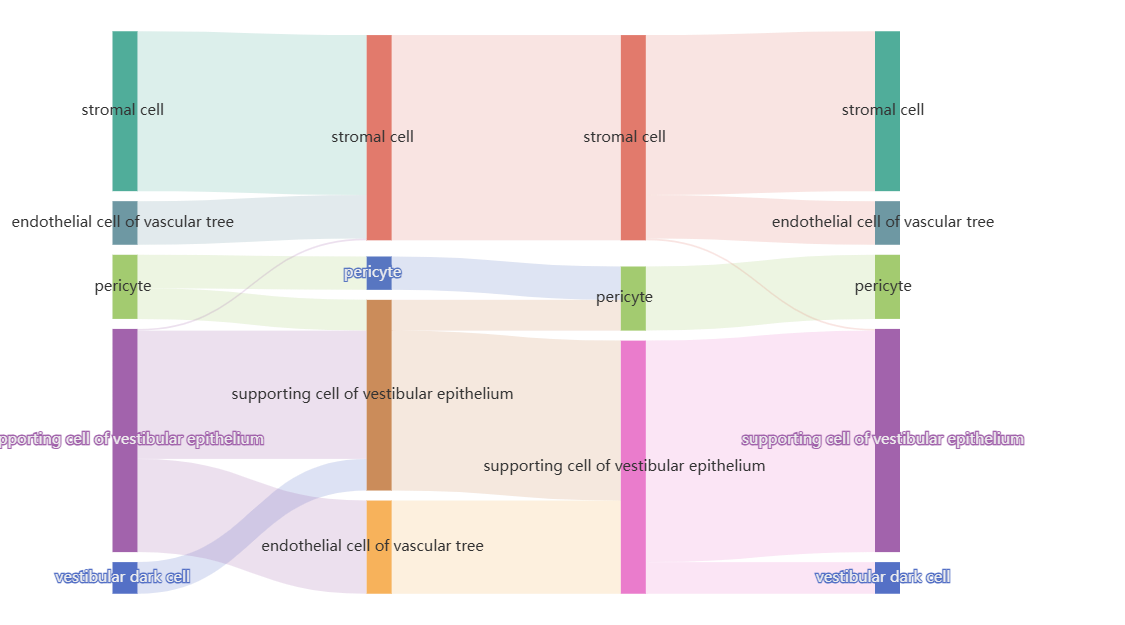}
\label{subfig:}
}
\subfloat[scMAE]{
\includegraphics[width=0.33\textwidth,keepaspectratio]{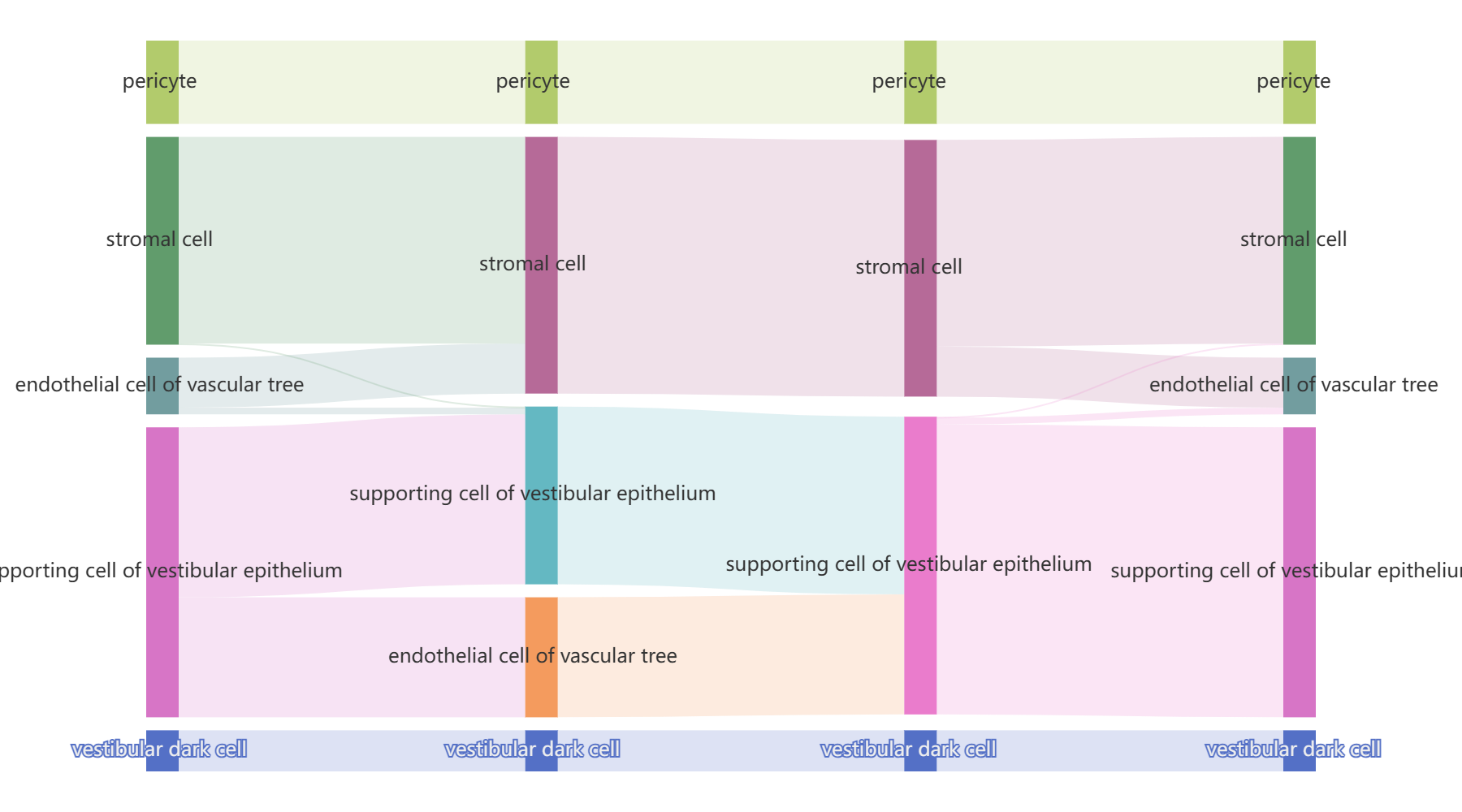}
\label{subfig:}
}
\subfloat[scCDC]{
\includegraphics[width=0.33\textwidth,keepaspectratio]{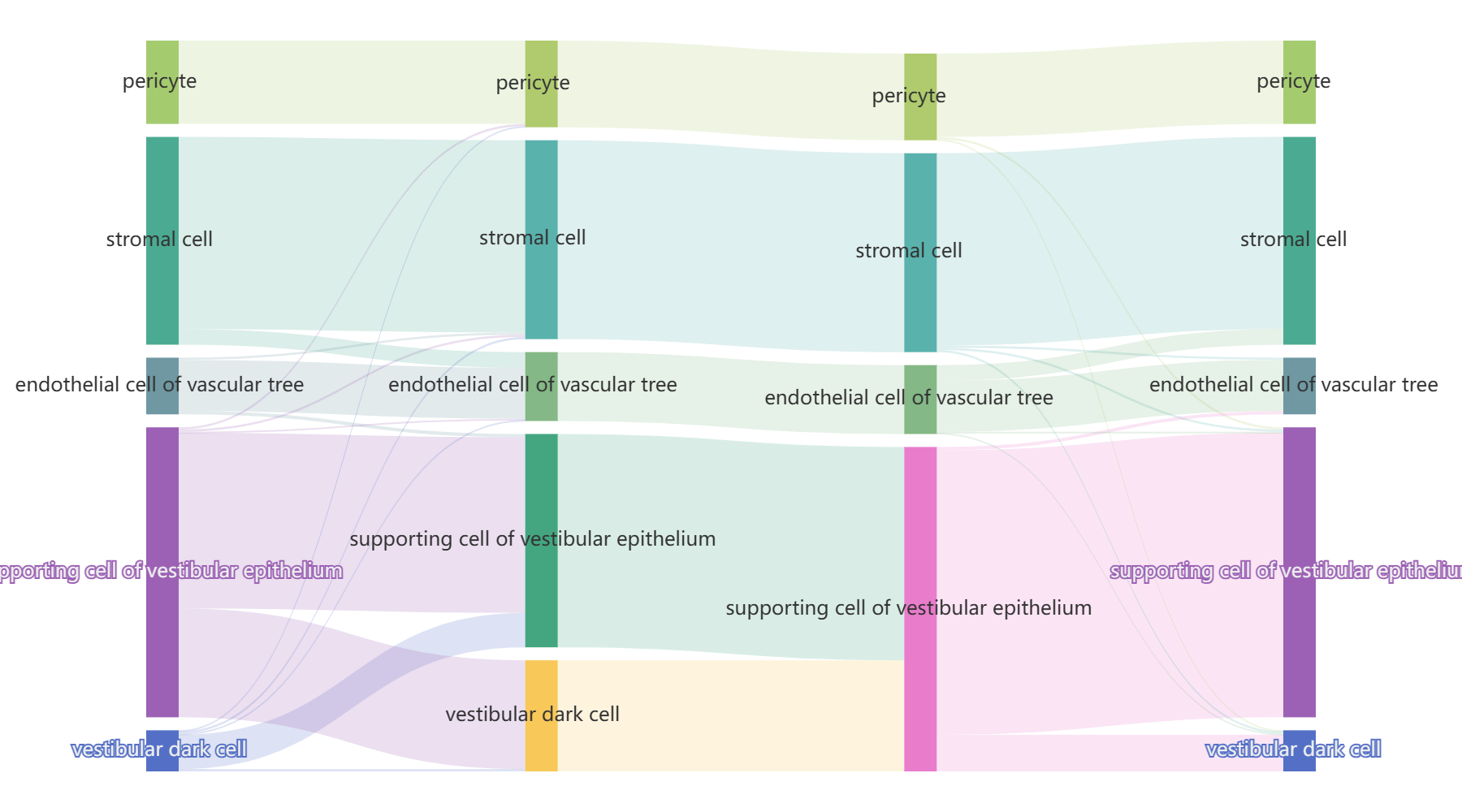}
\label{subfig:}
}
\caption{Case study on the \textit{Sapiens Ear Utricle}, summarizing the results of Leiden, scMAE, and scCDCG models, including representation learning, two-dimensional visualization, and marker gene-based cell type annotation.(m)-(o) presents four columns from left to right: Gold-standard labels, results of the Best-mapping annotation, results of the Marker-overlap annotation, and the Gold-standard labels.}
\label{fig:all_sapies}
\end{figure*}

The results, summarized in Fig.~\ref{fig:all_muris} for the \textit{Muris Limb Muscle} dataset and Fig.~\ref{fig:all_sapies} for the \textit{Sapiens Ear Utricle} dataset, consistently demonstrate that the standardized and high-quality data in~\methodname~enable reliable, reproducible analyses and support diverse analytical paradigms. This unified and versatile resource thus provides a solid foundation for a wide range of downstream single-cell tasks, from representation learning to cell type annotation, facilitating both methodological evaluation and biologically driven discovery.

\section{Conclusion}
We present~\textbf{\methodname}, an AI-ready standardized resource for single-cell RNA sequencing analysis that consolidates diverse biological datasets into a unified, analysis-ready framework.
By providing uniform preprocessing, standardized formatting, and multi-level annotations,~\methodname~addresses key challenges in reproducibility, comparability, and cross-dataset evaluation.
Importantly, \methodname~offers high-quality, standardized datasets that support a broad range of downstream analyses, including but not limited to clustering, classification, marker gene identification, and cell-type annotation, serving as a foundational resource to facilitate AI-driven model development and computational analyses in single-cell biology.
Looking forward, we aim to expand~\methodname~to include more species, additional tissue types, and complementary omics data, further strengthening its utility for single-cell research and AI-driven model development.

% \section*{Acknowledgment}

% The preferred spelling of the word ``acknowledgment'' in America is without 
% an ``e'' after the ``g''. Avoid the stilted expression ``one of us (R. B. 
% G.) thanks $\ldots$''. Instead, try ``R. B. G. thanks$\ldots$''. Put sponsor 
% acknowledgments in the unnumbered footnote on the first page.

\clearpage
\bibliographystyle{IEEEtran}
\bibliography{ref}
% \newpage 
% \input{7_appendix}

\end{document}